\documentclass[default]{sn-jnl}% Default
\usepackage{amssymb}
\usepackage[linewidth=1.5pt,linecolor=gray]{mdframed}

\newcommand\ahalf{a_{\rm 1/2}}
\renewcommand\d{{\rm d}}

\newcommand{\gbar}{g_{\rm bar}}
\newcommand{\gobs}{g_{\rm obs}}
\newcommand\Jv{{\boldsymbol J}}
\newcommand\kpc{\,{\rm kpc}}
\newcommand\kms{\,{\rm km\,s^{-1}}}
\newcommand\Lv{{\boldsymbol L}}
\newcommand{\Lsun}{L_{\odot}}
\newcommand{\Msun}{M_{\odot}}
\newcommand\Mstar{M_\star}
\newcommand{\MDM}{M_{\rm DM}}
\newcommand{\Mdyn}{M_{\rm dyn}}
\newcommand{\Mgas}{M_{\rm gas}}
\newcommand\rhalf{r_{\rm 1/2}}
\newcommand\pc{\,{\rm pc}}
\newcommand\Rhalf{R_{\rm 1/2}}
\newcommand\xv{{\boldsymbol x}}
\newcommand\vv{{\boldsymbol v}}

\newcommand\rhostar{\rho_\star}
\newcommand\rhogas{\rho_{\rm gas}}
\newcommand{\rhoDM}{\rho_{\rm DM}}
\newcommand{\rhotot}{\rho_{\rm tot}}

\newcommand{\Phitot}{\Phi_{\rm tot}}
\newcommand{\sigmalos}{\sigma_{\rm los}}
\newcommand{\sigmalosmax}{\sigma_{\rm los,max}}

\jyear{2021}%

\theoremstyle{thmstyleone}%
\theoremstyle{thmstyletwo}%

\theoremstyle{thmstylethree}%

\begin{document}

\title[Stellar dynamics and dark matter in dwarf galaxies]{Stellar dynamics and dark matter in Local Group dwarf galaxies}


\author*[1,2]{\fnm{Giuseppina} \sur{Battaglia}}\email{gbattaglia@iac.es}
\equalcont{These authors contributed equally to this work.}

\author[3]{\fnm{Carlo} \sur{Nipoti}}\email{carlo.nipoti@unibo.it}
\equalcont{These authors contributed equally to this work.}

\affil*[1]{\orgname{Instituto de Astrofísica de Canarias}, \orgaddress{\street{calle Via Lactea s/n}, \city{San Cristobal de La Laguna}, \postcode{E-38205}, \country{Spain}}}

\affil[2]{\orgdiv{Department of Astrophysics}, \orgname{University of La Laguna}, \orgaddress{\city{San Cristobal de La Laguna}, \postcode{E-38206}, \country{Spain}}}

\affil[3]{\orgdiv{Dipartimento di Fisica e Astronomia “Augusto Righi"}, \orgname{Università di Bologna}, \orgaddress{\street{via Gobetti 93/2}, \city{Bologna}, \postcode{I-40129},  \country{Italy}}}


\abstract{
When interpreted within the standard framework of Newtonian gravity and dynamics, the kinematics of stars and gas in dwarf galaxies reveals that most of these systems are completely dominated by their dark matter halos. These dwarf galaxies are thus among the best astrophysical laboratories to study the structure of dark halos and the nature of dark matter. We review the properties of the dwarf galaxies of the Local Group from the point of view of stellar dynamics. After describing the observed kinematics of their stellar components and providing an overview of the dynamical modelling techniques, we look into the dark matter content and distribution of these galaxies, as inferred from the combination of observed data and dynamical models. We also briefly touch upon the prospects of using nearby dwarf galaxies as targets for indirect detection of dark matter via annihilation or decay emission.} 

\date{Draft, December 14, 2021}

\maketitle

\today

\section{Introduction}\label{sec:intro}

Studies of the galaxy luminosity function indicate that most of the galaxies in the Universe are dwarfs, i.e.\ systems at least one order of magnitude less luminous than the Milky Way (MW). This finding, combined with the fact that, according to the standard cosmological model, structure formation occurs bottom-up (smaller galaxies form first), makes dwarf galaxies (DGs) extremely interesting systems in the context of galaxy formation and evolution studies (e.g.\ \cite{Cimatti2019}). Within the Newtonian framework, it is now well established that DGs tend to be dark matter (DM) dominated, so they are are among the best astrophysical systems to be used as laboratories to study the nature of DM.

Nowadays, the most popular hypothesis on the nature of DM is that it is composed of cold dark matter (CDM) particles (with masses $\gtrsim 1$ GeV) that, in essence, interact among themselves and with other particles only gravitationally (and are thus said, in jargon, to be weakly interacting). 
However, there are several alternatives that are not excluded, such as, for instance, models in which the DM consists of weakly interacting warm dark matter particles (with masses $\sim$ keV) or models in which the DM is self-interacting, i.e.\ made of particles that interact among themselves not only gravitationally (e.g.\ \cite{Bertone2018}).  
For a given DM particle model, DM-only cosmological simulations allow to predict robustly the properties of the DM halos, but these properties can be significantly altered by the presence of baryons, whose effects are difficult to predict in detail, due to the complexity of baryonic physics. The more a system is DM dominated, the less important are the effects of baryons on the properties of dark halos. Thus, the halos of the most DM dominated DGs are expected to be mainly the product of DM physics, and can be used to constrain DM particle models.
In fact, within the standard CDM framework, there is tension between the predictions of DM-only cosmological simulations and the observational data on the small scales of DGs (most notably, the so-called cusp/core, missing satellites, and too-big-to-fail problems). It is a matter of debate whether these problems can be solved within the CDM paradigm when the effect of baryons are properly accounted for or whether their solution requires alternative DM models. We refer the reader to \cite{Bullock2017} for a detailed discussion of these small-scale problems and of possible solutions.

The knowledge of the amount and distribution of DM in DGs is of great interest also to test galaxy formation models, because the hierarchical assembly of DM halos is believed to be the backbone of the process of structure formation. In particular, on the scales of DGs, the interplay between the gravitational potential well of the halo and stellar feedback is crucial to determine 
the star formation efficiency and thus the properties of the galaxy stellar component (\cite{Dekel1986}; see also, e.g., \cite{Arora2021} and references therein).
In this context, the lowest-luminosity DGs are especially interesting, because the masses of their DM halos are close to the critical masses below which galaxy formation is theoretically expected to be strongly inhibited by the effects of reionization UV feedback and inefficient radiative cooling (e.g.\ section 10.5 of \citep{Cimatti2019}).

Given that DGs are intrinsically faint, their stellar dynamics and DM mass distribution can be studied in detail only if they are sufficiently nearby. For this reason in this review we  consider only DGs belonging to the Local Group (LG), i.e.\ the almost one hundred known DGs within $\approx 1$ Mpc from the Milky Way, whose stars are spatially resolved with currently available observational facilities.
In particular, we review the observed internal stellar kinematic properties  of LG DGs and what they imply for inferences on the amount and distribution of their DM. Given that the great majority of the LG DGs is devoid of gas, our focus is on stellar dynamics, but, of course, when a gaseous disc is present, its rotation curve can be used as an additional independent tracer of the galaxy gravitational field (e.g.\ \cite{Read2017a}; \cite{Leung2021}; see also \cite{Lelli2022}).

We adopt the nomenclature by \cite{Simon2019} for the gas-free and passively evolving systems and refer to them as dwarf spheroidals (dSphs) or ultra-faint dwarfs (UFDs) when, respectively, brighter or fainter than absolute $V$-band magnitude $M_V= -7.7$ (as a consequence of their challenging detectability, the majority of known UFDs are around the MW).
For DGs containing gas, we refer to them as dwarf irregulars (dIrrs) if currently forming stars or transition types (dTs) if without current star formation. For historical reasons we still use the term “classical"
DGs to indicate systems known before the advent of the Sloan Digital Sky Survey. We will classify as “isolated" those DGs with a present-day distance of at least 
400 kpc from the closest large LG spiral. We base this choice on an empirical indication of environmental effects on the observed properties of LG DGs, known as the morphology-density relation \citep[see e.g. review by][]{vandenbergh1999}, i.e. the fact that the great majority of the LG DGs devoid of gas are clustered around the MW or M31, while the great majority of those with HI detections are found at more than 400 kpc from the Milky Way or M31’s center \citep[see][for the most recent compilation of HI measurements]{Putman2021}.
 We exclude from our analysis the Large Magellanic Cloud (LMC), the Small Magellanic Cloud, M~32 and M~33, due to their clearly different characteristics with respect to the rest of the population, as well as the heavily tidally disrupted Sagittarius dSph. 

Throughout the paper we will be working in the underlying assumption of Newtonian gravity and dynamics. However, we recall that alternative frameworks do exist, most notably the Modified Newtonian Dynamics (MOND; \cite{Milgrom1983}), which appears able to reproduce the observed stellar velocity dispersion profiles of most classical dSphs \cite{Angus2008,Hernandez2010,Serra2010}, but has problems with most UFDs \citep{Safarzadeh2021}. A remarkable exception is the UFD Crater~II, for which the stellar velocity dispersion predicted by \cite{McGaugh2016a} assuming MOND has been later confirmed by observations \citep{Caldwell2017}.

This is not the first time that the internal kinematic properties of LG DGs are covered in a review since the first of such works by \cite{Mateo1998}, albeit with different focus: \cite{Tolstoy2009} treated mostly the observational aspects, \cite{Battaglia2013, Walker2013, Strigari2018b} covered the observed internal kinematics, methods of dynamical modelling and implications for DM determinations of MW satellites, with the former concentrating on the classical dSphs and the latter two including UFDs; recently, \cite{Simon2019} reviewed in detail the properties of UFDs. We will therefore try to keep our article self-contained, but build on these earlier works and highlight results posterior to most of these reviews. 
 
 \section{Observed internal stellar kinematics}
 \label{sec:obskin}
 
 Ideally, a complete definition of the dynamical properties of a system requires 6D phase-space information of large numbers of stars. In practice, for the systems considered here, the relative precision attainable for distances to the individual stars is, at present, typically of the order of or larger than the size of the dwarf galaxy itself, even when using precise distance indicators such as RRLyrae variable stars \citep[see e.g.][]{Muraveva2020}. 
Therefore, in general, the available quantities  are two spatial coordinates (the components of the position vector in the plane of the sky) and from zero up to three components of the velocity vector (l.o.s.\ velocity and two proper motions, $\mu_{\alpha,*}$ and $\mu_{\delta}$). Since the advent of {\it Gaia} DR2, in several LG DGs there are more stars with a measurement of the proper motion than of the l.o.s.\ velocity \citep[e.g. see][]{Fritz2018, McConnachie2020, Battaglia2021}. Nonetheless, the uncertainty in the transverse velocity is far larger than that on the l.o.s.\ component: taking red giant branch (RGB) stars around 1 mag below the tip of the RGB in Draco as a reference, the uncertainty on the {\it Gaia} eDR3 $\mu_{\alpha,*}$ would translate into an uncertainty in transverse velocity around 30 km s$^{-1}$, while l.o.s. velocities are measured to better than $\pm$1-2 km s$^{-1}$ \citep[using as guideline the spectroscopic sample of][]{Walker2015}. It is therefore still on the l.o.s.\ component of the velocity vector that dynamical modelling rests upon.
 
In this section we emphasise those observational aspects that form the main input to the dynamical modelling or serve to verify some of the assumptions, i.e.\ ordered and random motions, the presence of distinct stellar components, kinematic peculiarities and contamination of kinematic samples by unresolved binaries and MW stars (see Box~1 for the latter). 

Most of the LG DGs are located in the vicinity of a much larger host and there is therefore an interest in identifying what systems were/are particularly affected by tidal interactions. It has been the case only recently that constraints on the orbital history of 
most MW satellites can be placed, albeit with heterogeneous levels of precision, thanks to the systemic proper motions measurements enabled by the second and early third data release of the {\it Gaia} mission [hereafter, {\it Gaia} DR2 and eDR3]  \citep[e.g.][]{Helmi2018, Simon2018, Fritz2018,  McConnachie2020, Battaglia2021, Li2021}; the situation is less favourable for M31 satellites. Since this is a still evolving area, in this review we do not cover this aspect specifically; suffice to say that considering MW satellites, there are a few systems that can reach within 30 kpc from the center of our Galaxy, which puts them at risk of significant tidal disturbances. Notably, it has become clear that specific assumptions on the gravitational potential of the MW (not only the assumed MW halo mass, but also whether the infall of a massive LMC is accounted for) has a strong impact on the inferred orbital parameters for a significant number of its satellite galaxies \citep[e.g.][]{Patel2020, Battaglia2021}.

\subsection{Ordered motions} \label{sec:ordered}
Velocity gradients in the l.o.s. and/or proper motion components of the velocity vector for stars in LG DGs can arise from a variety of causes, such as intrinsic rotation, streaming motions and/or tidal disruption. The search for such velocity gradients has therefore a fairly long history, given that these are a manifestation of processes related to the formation and evolution of the system and that can impact choices made in the dynamical modelling (e.g.\ the use of non-rotating models or the assumption of equilibrium). Examples of velocity gradients most likely induced by tidal effects can be seen in NGC~205 \citep{Geha2006}, Tucana~III \citep{Li2018} and Antlia~II \citep{Ji2021} (orbit integration enabled by {\it Gaia} proper motions clearly supports this interpretation for Tucana~III, while for Antlia~II there is a dependence on the assumed MW gravitational potential); on the other hand, velocity gradients detected in galaxies where the time elapsed from the last pericenter passage exceeds a few tens of crossing times are unlikely to be due to tidal disturbances, since the galaxies would be expected to have reached a new equilibrium configuration on those timescales \citep[e.g.][]{Penarrubia2009}.  For objects with large angular sizes, such as MW satellites, also the relative motion of the DG with respect to the Sun can cause a measurable velocity gradient \citep[“perspective rotation", see e.g.][]{vanderMarel2001}, but this can be accounted for if the bulk motion of the system is precisely known.

Independently of the mechanism causing the presence of a velocity gradient, in LG DGs these are expected to be of a fairly low amplitude and increase slowly 
as a function of the distance from the galaxy's centre; this implies that data-sets focusing on the central regions will not be particularly sensitive to the detection of such features. Currently, the sample sizes, velocity precision and spatial coverage of the data are often insufficient to pin down the amplitude and direction of these velocity gradients, but useful limits can be placed. 

Thus far, the overwhelming majority of searches for velocity gradients has been carried out on the l.o.s.\ component of the velocity vector, with most works concentrating on one or few systems \citep[e.g.][]{Lewis2007, Battaglia2008, Walker2008, Fraternali2009, Kirby2014}.  \cite{Wheeler2017} performed a comprehensive analysis of literature data-sets for 40 LG DGs, up to stellar masses $\sim 10^{8}$M$_{\odot}$.
 Moderate or strong evidence for velocity gradients was found in only 7 out of 40 galaxies. Five of these 7 galaxies are isolated, so their gradients are imputable to internal processes. The measured $V/\sigma$ is $\le2$ for the whole sample and $\lesssim 1$ for more than 3/4 of it. Assuming the velocity gradients are caused by rotation, these findings would imply that rotation is typically sub-dominant with respect to random motions. 
In general, a model with radially increasing rotation speed is favoured by the data over a model with constant rotation speed, confirming the importance of spatially extended data-sets to detect these features.

Since the work by \cite{Wheeler2017}, searches were  performed in new spectroscopic data-sets for Leo~V \citep{Collins2017}, Leo~A, Aquarius, Sag dIrr \citep{Kirby2017}, Phoenix \citep{Kacharov2017}, Leo~II \citep{Spencer2017}, Cetus \citep{Taibi2018}, And~XIX \citep{Collins2020}, Aquarius \citep{Hermosa2020}, Tucana \citep{Taibi2020}, NGC~6822 \citep{Belland2020}, Hercules \citep{Gregory2020}, And~I, And~III, And~V, And~VII, And~X \citep{Kirby2020}. 
Statistically significant velocity gradients and/or with at least a weak/moderate preference over a non-rotating model were found only in Phoenix, Aquarius, NGC~6822 and IC~1613, with a $V/\sigma \sim 1$ for the first two galaxies and $<$1 for the latter two (see Sect.~\ref{sec:peculiar} for peculiarities in some of these DGs).  $V/\sigma$ is not necessarily constant with radius: for instance WLM has $V/\sigma$ = 2.5 in the last measured point, to be compared to the global value of 1.1 (\cite{Leaman2012}; see also the case of NGC~147 \cite{Geha2010}).

To date, the only measurement of ordered motions onto the plane of the sky was performed by \cite{Martinez2021} with {\it Gaia} eDR3. Even though transverse velocities with uncertainties up to 100 km s$^{-1}$ were used, a global internal transverse velocity was detected in Sculptor, Fornax and Carina
(at the 3, 2, and 2 $\sigma$ level, respectively), and glimpses of its spatial variation could be gathered for Ursa Minor, Fornax and Sculptor. In all cases, the ratio of ordered versus random motion is less than 0.5; Carina displays a larger ratio, but at a low significance, V  /$\sigma =  1.45 \pm 0.73$. 

Overall, ordered motions are subdominant with respect to random motions in 
the stellar kinematics of LG DGs, except in a handful of cases.

\subsection{Random motions}

\begin{figure}[t!]
\centering
\includegraphics[width=\textwidth]{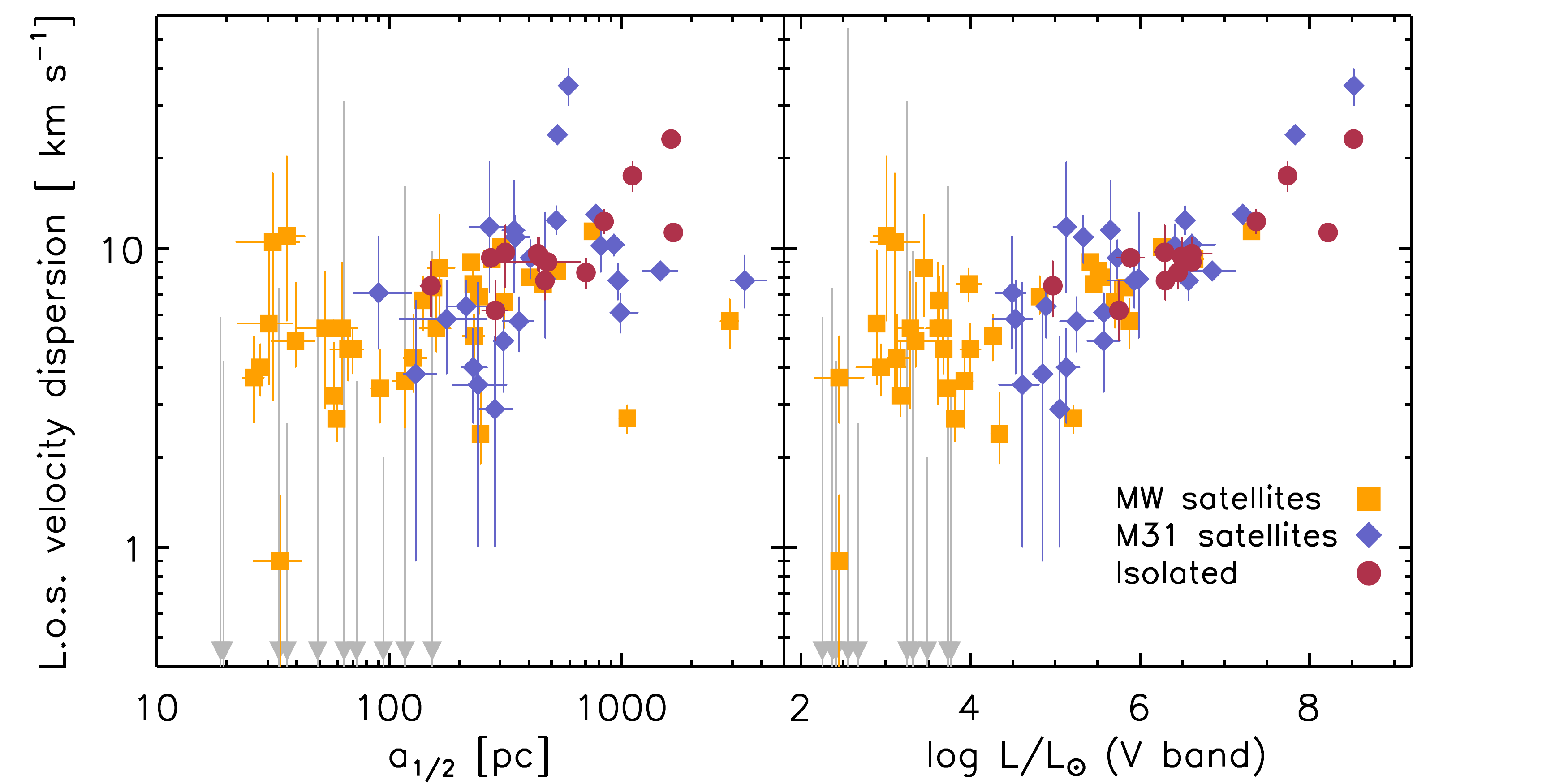}
\caption{
L.o.s.\ stellar velocity dispersion as a function of semi-major axis of the half-light ellipse (left) and luminosity in the $V$-band (right) for the 81 LG DGs listed in Tables~\ref{tab:mdyn} and \ref{tab:unresolved}, distinguishing satellites of the MW (squares), satellites of M 31 (diamonds) and isolated galaxies (circles), see Section~\ref{sec:intro}.
The error bars are the 16-th and 84-th confidence limits. The grey downward arrows indicate upper limits for the galaxies with an unresolved value of $\sigmalos$, which are all MW satellites (see Tab.~\ref{tab:unresolved} for more details).
}
\label{fig:sigma}
\end{figure}

As discussed in Sect.~\ref{sec:ordered}, the internal stellar kinematics of most LG DGs is dominated by random motions and it is therefore mainly from these that we infer information on the DM mass and distribution of these galaxies.  

At present, only a global value of the stellar l.o.s.\ velocity dispersion, $\sigma_{\rm los}$, is available for the vast majority of LG DGs (see Fig.~\ref{fig:sigma} for an overview of $\sigma_{\rm los}$ as a function of semi-major axis of the half-light ellipse $a_{1/2}$ and $V$-band luminosity $L$). Knowledge of the shape of the l.o.s.\ velocity dispersion profile is mostly limited to the MW classical dSphs, some bright M31 satellites and a handful of isolated LG dwarfs. The uncertainties on the UFDs $\sigma_{\rm los}$ remain large, in particular for those fainter than $M_V \sim -4.5$ ($L \sim 5\times 10^3$ $L_{\odot}$). For example, if we take as representative a value of 5 km s$^{-1}$ for $\sigmalos$, the typical relative uncertainty on $\sigmalos$ is $\sim$20\% at $\log(L/L_{\odot}) \sim 4.5$ and $\sim$40-60\% at $\log(L/L_{\odot}) \sim 3.5$.

No statistically significant distinction is seen between MW dSphs, M31 dSphs and isolated LG DGs of various types \citep{Tollerud2012,Kirby2014,Higgs2021}, nor between dwarfs in and off the great plane of Andromeda satellites \citep{Collins2015}, in that e.g. they occupy similar regions of the $\sigma_{\rm los}$ versus size plane. 
A clear exception is represented by Crater~II, with a $\sigma_{\rm los} \sim 2.5 \kms$ and  $a_{1/2} \sim 1000$ pc \citep{Caldwell2017, Fu2019, Ji2021}, to be compared to the 8-10 $\kms$ of similarly extended DGs: its low $\sigma_{\rm los}$ might be the result of repeated tidal stripping.
However, it should be considered that, as it is the case for the UFDs, there is a certain degree of heterogeneity in the size and quality of the spectroscopic samples for isolated and M31 dSphs: more homogeneous spectroscopic data-sets are needed to confirm that
MW dSphs, M31 dSphs and isolated LG DGs have similar properties. In particular, accurate determinations of velocity uncertainties are important for systems with low intrinsic $\sigmalos$ as LG DGs and these are often difficult to obtain, especially for low signal-to-noise ratio spectra, hence it is also possible that some estimates of $\sigmalos$ will be revised.  For example, an early study on Cetus reported a large $\sigma_{\rm los} =  17\pm 2$ km s$^{-1}$ \citep{Lewis2007}, while more recently the value has been converging to 8-11 km s$^{-1}$ \citep{Kirby2014, Taibi2018}; similarly, Tucana $\sigma_{\rm los}$ had been measured to be $\sim  15\pm 3$ km s$^{-1}$ 
\citep{Fraternali2009, Gregory2019},
while recent analyses place it at a most likely value of $\sim$6 km s$^{-1}$ \citep{Taibi2020}. 

 As for the MW classical dSphs, we refer the reader to the review articles by \cite{Battaglia2013, Walker2013} for detailed discussions on their l.o.s.\ velocity dispersion profiles. Here we highlight that these systems enjoy sizeable spectroscopic samples of accurate l.o.s.\ velocities (from $\sim$200 to $\sim$2500 probable member stars), 
 which reach out to their outskirts;  
 this clearly allows for a very different level of detail in the dynamical modelling with respect to the cases above (see Section~\ref{sec:dynmod}). The $\sigmalos$ profiles are approximately constant with radius, with mild declines or increases observed in some cases \citep[e.g.\ figure~12 of][]{Walker2013}. A common feature is the presence of wiggles/bumps, which do not appear to be consistent with statistical fluctuations, at least on the basis of the published uncertainties. Some systems at the bright end of the LG DG luminosity function, like NGC~147 \citep{Geha2010} and WLM \citep{Leaman2012}, exhibit declining $\sigmalos$ profiles accompanied by rotational velocities increasing as a function of radius, while for others the trend with radius appears less clearly defined. 

Thus far we discussed 1D profiles obtained by binning the data along concentric (circular or elliptical) annuli in the radial direction. There are however indications that the amplitude and shape of the $\sigmalos$ radial profile might vary as a function of azimuthal angle \citep[e.g.][]{Zhu2016, Hagen2019, Hayashi2020}, hence data-sets mapping the kinematics in 2D could in principle be an asset for the dynamical modelling, due to the larger number of observational constraints to be reproduced. 

One of the most exciting developments of the last few years, from the point of view of observations of the internal kinematics of LG DGs,  were the very first measurements of random motions from the components of the velocity vector along the plane of the sky \citep[][for Sculptor and Draco, respectively]{Massari2018, Massari2020}. This achievement was made possible by a synergistic use of {\it Gaia} 
and HST data, which led to an impressive reduction in the proper motion uncertainties with respect to the use of {\it Gaia}-only data (e.g. one order of magnitude at $G \sim 19.5$ for the case of Draco, where the {\it Gaia-}DR2 only uncertainty per proper motion component was $\sim$0.5 mas yr$^{-1}$). The measurement of the internal random motions on the plane-of-the-sky has allowed the first observational quantification of the ratio of random motions in the different components of the velocity vector, a key ingredient in several modelling techniques (see Sect.~\ref{sec:dynmod}). While the uncertainties are presently still too large to disentangle between competing models, this gives a first taste of what future astrometric measurements can provide (see Sect.~\ref{sec:outlook}).

\subsection{Chemo-kinematics} \label{sec:chemokin}

The availability of large and spatially extended samples of individual stars with l.o.s.\ velocity and metallicity or age information for some of the classical LG DGs has allowed detailed joint analyses of their kinematic and chemical properties (“chemo-kinematics", or we might say “chrono-kinematics" if age information is used), unveiling a rather surprising level of complexity. Sculptor, Fornax, Sextans, Carina and Ursa Minor are known to host multiple chemo-kinematic components (CKCs), i.e.\ their stellar population can be described as the super-position of components of different mean metallicity, spatial distribution, and kinematics \citep[e.g.][]{Tolstoy2004, Battaglia2006, Battaglia2011, Walker2011, Amorisco2012, Fabrizio2016,Kordopatis2016, Pace2020}; hints are found in Leo~II  \citep{Spencer2017}, in the isolated Cetus and Tucana dSphs \citep[][respectively]{Taibi2018, Taibi2020}, and there have been claims also for Canes Venatici~I \citep[][]{Ibata2006}
and Bootes~I \citep{Koposov2011}. Not for all systems it is clear, or it has been explored, whether more than two components are present; so far, three CKCs have been reported for Carina and Fornax (and there might be hints in Ursa Minor), 
while Sculptor appears naturally described by two components, as supported by a bimodal non-parametric distribution function in energy and angular momentum space 
\citep{Breddels2014}. Photometric data show that spatial variations in the distribution of stellar ages or metallicity are common in LG DGs; it would therefore not be surprising if future spectroscopic studies with larger samples unveil multiple CKCs in more systems.

In all cases, with the exception of Carina, there is a well defined ordering, in the sense that the more metal-rich stars have a lower velocity dispersion than the metal-poor stars and are more centrally concentrated. 
The super-position of these chemo-kinematic components might be responsible for some of the bumps/wiggles observed in the $\sigmalos$ profiles, as suggested by \citep{McConnachie2007}.

The presence of multiple CKCs has triggered a new avenue for the dynamical modelling (Sect.~\ref{sec:dynmod}), because CKCs are independent tracers of the DG gravitational potential. It is therefore important to determine the properties of these CKCs, since they might differ from those of the overall stellar population and 
from each other, or significantly deviate from common assumptions made in the modelling (as those of sphericity, lack or rotation or flatness of the $\sigmalos$ profiles). For example, a declining $\sigmalos$ profile has been detected for one of the CKCs in some systems (Sculptor: \cite{Battaglia2008}; Fornax: \cite{Amorisco2012Fnx}; Ursa Minor \cite{Pace2020}). In the massive LG DGs WLM and NGC6822, whose stellar component is rotating, the amount of rotational support is found to vary with age \citep{Leaman2012} or metallicity \citep{Belland2020}. Interestingly, there are also a couple of known cases of multiple stellar components of strikingly different ellipticities: e.g. in Ursa Minor the metal-rich component is significantly more flattened (ellipticity $\epsilon = 0.74 \pm 0.04$) than the metal-poor one ($\epsilon= 0.33_{-0.11}^{+0.09}$); 
see also the case of Sextans \citep{Cicuendez2018}.

\subsection{Peculiar kinematic properties} \label{sec:peculiar}

A few of the LG dwarf galaxies that had had the benefit of being studied in detail with large spectroscopic samples display peculiar kinematic properties, such as rotation misaligned with the major axis of the stellar component and/or kinematic substructures on a large scale. 

For example, statistically significant prolate rotation, i.e.\ rotation around the major axis, has been detected in And~II \citep{Ho2012} and in Phoenix \citep{Kacharov2017}, while hints are found in Ursa Minor \citep{Pace2020} and in NGC6822 \citep[e.g.][]{Belland2020}. These systems also show other complexities. The major axis of the spatial distribution of the youngest stars is orthogonal to that of the bulk of the population in Phoenix \citep[e.g.][]{Hidalgo2009, Battaglia2012} and it is strongly misaligned in NGC 6822 \citep{Thompson2016}; in both cases, it is aligned with the direction of maximum rotation. In And~II, \cite{delPino2017AndII} found the rotational properties to change with metallicity/age, in that the metal-poor/old stars do exhibit prolate rotation, but the metal-rich/intermediate-age ones do not. 
Also Fornax and Sextans are known to display unexpected kinematic behaviours; for example, for Fornax, early detections of a double-peaked l.o.s.\ velocity distribution for the metal-poor stars \citep{Battaglia2006} have been put into better focus with findings of different directions of the angular momentum vector/rotating patterns as a function of metallicity \citep{Amorisco2012Fnx, delPino2017Fnx} (see \cite[][]{Cicuendez2018, Kim2019} for Sextans).

Mergers are often invoked to explain the observed peculiar kinematics.  This appears to be a likely explanation for the cases of prolate rotation, as both idealized and cosmological simulations find a connection between the timing of the appearance of prolate rotation and the occurrence of significant mergers events, both at large masses and on the scale of DGs (see e.g.\ \cite{Lokas2014, Cardona-Barrero2021}). In And~II, the detection of a stellar stream with stellar population properties compatible with being a disrupted smaller DG seemed to confirm this hypothesis \citep{Amorisco2014}, although such detection is controversial \citep{delPino2017AndII}.

Understanding whether mergers are responsible for the observed peculiar kinematic properties would have an intrinsic interest in validating the hierarchical galaxy formation scenario down to the smallest galactic scales, where direct signs of mergers are available only down to the scale of LMC-like galaxies \citep[e.g.][]{Annibali2016}. This would require finding in cosmological hydrodinamic simulations
merger configurations able to reproduce both the peculiarities observed in the various LG DGs and the frequency of their occurrence in a quantitative way. Concerning the topic of this review, i.e.\ the use of the observed kinematic properties of LG DGs to infer their DM content and distribution, we argue it would be additionally necessary to evaluate whether the impact on the internal dynamics of the remnant might put into question the application of stationary dynamical models. 
 
\bigskip
% {\bf Start-Box-1}
\begin{mdframed}

\subsection*{Box 1. Contamination of kinematic samples} \label{sec:binaries}
\bigskip

Two types of 'contaminants' can be present in a kinematic sample of a DG: stars that do not belong to the DG and stars whose velocity can be considered as spurious, because it contains a contribution unrelated to the orbit of the star within the DG, as it is the case for unresolved binaries.

The observed kinematic samples are in general contaminated by field stars of the MW (and of M31, for M31 satellite DGs) that happen to lie on the same l.o.s.\ as the DG.  Contamination can be addressed by either cleaning the sample, i.e.\ removing non-member stars based on hard-cuts, by assigning a probability of membership \citep[e.g.][]{Battaglia2008, Walker2009c}, or by adding a suitable population of contaminants to the model \citep[][]{Zhu2016,Pascale2018, Horigome2020}.
Several characteristics can be used to identify non-members, e.g.\ l.o.s.\ velocity, location on the colour-magnitude diagram, metallicity or spectroscopic indicators of surface gravity \citep[see e.g.][for a discussion]{Battaglia2013}, and of course nowadays also parallaxes and proper motions. Overall, it appears that residual contamination in spectroscopic samples does not significantly affect the global values of $\sigmalos$, even for UFDs \citep[see][]{Simon2019}.
It is likely that the use of {\it Gaia} astrometry will be helpful in better pinning down the shape of the $\sigmalos$ profile in the outskirts of bright MW satellites. 

The question of whether the observed $\sigmalos$ of LG DGs might be inflated by unresolved binaries has naturally arisen since the first studies of their internal kinematic properties \citep[see discussion in][and references therein]{Battaglia2013, Walker2013, Simon2019}. To date, answering this question has been rather tricky, given that the benefit of multi-epoch observations has been restricted to only sub-sets of the stars with spectroscopic observations, if any.

There is widespread agreement that the measured values of $\sigmalos$ should not be significantly inflated for systems with fairly large intrinsic $\sigmalos$. On the other hand, a large impact is possible in those DGs with a low intrinsic $\sigmalos$ \citep[][]{McConnachie2010, Dabringhausen2016, Spencer2017}:
for instance, simulations by \cite{Spencer2017}
(see their figure~11) suggest that a binary fraction of 0.5 can potentially lead to an increment of $\sim$15\% and 80\% in the observed $\sigmalos$ from an intrinsic $\sigmalos$ of 6 and 2 km s$^{-1}$, respectively; the inflation becomes even more dramatic for lower values of the true dispersion.
This refers to samples of RGB stars, but the problem could be exacerbated for samples of stars of higher surface gravity, which, with their smaller radii, could form part of tighter binary systems.
In general, the velocity dispersion obtained by removing stars exhibiting statistically significant radial velocity variations is similar to that obtained from the whole sample of members, except in a few UFDs,  especially in cases of small samples \citep[see][]{Simon2019}. A different approach to quantify the possible impact of binaries onto $\sigmalos$ is to model the contribution of binary orbital motions when analysing the samples of l.o.s.\ velocities, as done in 
\cite{Martinez2011, Minor2019}
for Segue~1 and Reticulum~II: these studies conclude that it is very unlikely that the stellar random motions in these UFDs can be entirely accounted for by binary orbital motions, but stress the need for multi-epoch observations, since such conclusion cannot be firmly reached with only single-epoch data.

Given the challenge of gathering large samples of accurate l.o.s.\ velocities from multi-epoch observations in sparsely populated systems as the UFDs, it is worth asking whether the more populous, but still metal-poor, classical MW dSphs could serve as “templates" for the binary star properties of UFDs. 
The analysis of the available data-sets  
\cite{Minor2013, Spencer2018} suggests that the binary population in Milky Way dSphs differ from each other in their binary fractions, period distributions, or both;
for example, \cite{Spencer2018} find that the binary fractions are spread over a range of values with a width of at least 0.3-0.4 (if the period distribution does not vary across systems) or, if the DGs analyzed share the same binary fraction, then they have different period distributions.
Hence at present these DGs cannot be used to predict exactly the effect of binary contamination of UFDs, but only to place rather broad limits. 

\end{mdframed}

%{\bf End-Box-1}

\section{Excursus on dynamical modelling}
\label{sec:dynmod}

As pointed out in Section~\ref{sec:intro}, one of the main aims of studying the stellar dynamics of DGs is  obtaining information on the mass density distribution of their DM halos. For this purpose one needs to estimate the galaxy gravitational potential, which can be inferred from the kinematics of the stars via dynamical models, usually based on the assumption that the galaxy is isolated and stationary.  In the case of satellite DGs, such as the dSphs of the LG, the validity of this  assumption must be checked on a case by case basis, because the effects of the host galaxy tidal field depend on the orbit as well as on the structural properties of the dwarf \citep{Bullock2005,Penarrubia2008,Errani2015,Errani2020,Nipoti2021}. 
For instance, the relative unimportance of {\em present-day} tidal effects has been ascertained using $N$-body simulations for Fornax \citep{Battaglia2015} and for Sculptor \citep{Iorio2019}, which justifies modelling these systems as isolated, even if they may have experienced significant tidal stripping in the past \citep{Genina2020b,Borukhovetskaya2021a}.

A DG, assumed stationary, can be modelled as a multi-component collisionless system, with stellar, gas and DM density distributions $\rhostar$, $\rhogas$ and $\rhoDM$, respectively, in equilibrium in the total gravitational potential $\Phitot$, given by $\nabla^2\Phitot =4\pi G\rhotot$, where $\rhotot=\rhostar+\rhogas+\rhoDM$. The stellar component is fully described by a time-independent distribution function (DF) $f(\xv,\vv)$, such that 
\begin{equation}
    \rhostar(\xv)=\int f(\xv,\vv)\d^3\vv,
\end{equation}
where  $\xv$ and $\vv$ are, respectively, the position and velocity vectors. 
In the presence of more than one stellar population (for instance a metal-poor and a metal-rich stellar population; Section~\ref{sec:chemokin}), one can model each of them as a different component with its DF $f_k$, such that $f=\sum_k f_k$.
In spherical symmetry, $\Phitot$ can be expressed in terms of the dynamical mass $\Mdyn(r)\equiv 4\pi\int_0^r\rhotot(r')r'^2\d r'$ within a sphere of radius $r$, which is related to $\Phitot(r)$ by
\begin{equation}
    \frac{\d \Phitot}{\d r}=\frac{G\Mdyn}{r^2}.
\end{equation}
For instance, for a spherical galaxy with no gas $\Mdyn(r)=\Mstar(r)+\MDM(r)$, where $\Mstar(r)$ and $\MDM(r)$ are, respectively, the stellar and DM mass profiles. Starting from the concept of DF, there are different strategies to apply it in practice to build a dynamical model. In Box~2 we list some of these approaches,  referring the reader to \cite{Binney2008} and \cite{Ciotti2021} for detailed treatments.

In order to infer information on the intrinsic properties of a  DG, different dynamical models  are compared with observed velocities and projected positions of stars  (Section~\ref{sec:obskin}).
Either binned or discrete data can be used as input for the dynamical models. Binned data usually comprise profiles of stellar surface number density, l.o.s.\ velocity dispersion, and higher moments of the velocity distribution (e.g.\ \cite{Lokas2005,Lokas2009,Strigari2010,Battaglia2013,Breddels2013a,Richardson2013}), complemented by
proper-motion velocity dispersion measurements, when available \citep[][]{Massari2018,Strigari2018a,Massari2020}. Discrete data are mainly used for the kinematics and consist in individual measurements of stellar velocities
that contribute to the likelihood of a model, depending on the velocity distribution predicted by the model at the position of the star \citep{Zhu2016,Pascale2018,Read2021}.
When distinct stellar populations are present, the input data can be considered separately for each population (\citep{Battaglia2008,Amorisco2012,Agnello2012,Strigari2017,Pascale2019}) or jointly \citep{Zhu2016}.

The output of the model-observation comparison usually consists in confidence regions in the space of the model parameters that allow us to constrain the intrinsic (non observable) properties of the DG.
For spherical models, the main outputs are $\Mdyn(r)$ and the profiles of the velocity dispersion tensor components (which contain information on the anisotropy of the velocity distribution). For axisymmetric models, one obtains also the profiles of intrinsic flattening and rotation speed, assuming or inferring the inclination with respect to the l.o.s..
The output of dynamical modelling can suffer from the so-called {\em mass-anisotropy degeneracy} \cite{Binney1982}, that is the fact that models with different combinations of  the stellar velocity and  total mass distributions can perform similarly, when compared with data. Such a degeneracy can be alleviated when discrete kinematic data or  profiles of higher velocity moments are used as input for the model-observation comparison (see \cite{Read2021} and references therein). 

\bigskip
%{\bf Start-Box-2}
\begin{mdframed}

\subsection*{Box 2. Dynamical modelling techniques} \label{sec:modtech}

Here we briefly describe the main dynamical modelling methods used to infer the intrinsic properties of galaxies from observations of their stellar components, reporting references to  works in which they have been applied to  DGs.

\begin{itemize}

\item {\em Models based on analytic DFs}. Exploiting Jeans' theorem, one assumes analytic forms for the DFs as functions of integrals of motions. These models are powerful and predictive, because the DF is known, and intelligible, because the DF is analytic.
DFs of the form $f=f(E)$, where $E$ is the orbital energy, which describe components with isotropic velocity distribution, have been applied to dSphs by \citep{Amorisco2011}. In spherical symmetry one can construct anisotropic components with DFs $f=f(E,L)$, where $L$ is the magnitude of the orbital angular momentum $\Lv$ \citep{Wilkinson2002,Amorisco2012,Strigari2017}. Axisymmetric components can be built with DFs $f=(E,L_z)$, where $L_z$ is the component of $\Lv$ along the symmetry axis. As an alternative to the classical integrals $E$ and $\Lv$, the action integrals $\Jv$ can be used as arguments of the DF \citep{Williams2015,Pascale2018,Pascale2019} to build spherical, axisymmetric or triaxial models. 
 
\item {\em Schwarzschild} \citep{Schwarzschild1979} {\em orbit superposition methods}. For  given $\Phitot$,  one constructs a library of numerically integrated orbits, which are then weighted and combined to reproduce $\rhostar$.  This method has been applied to build both spherical \citep{Breddels2013a,Breddels2013b,Kowalczyk2017,Kowalczyk2018,Kowalczyk2019} and axisymmetric  \citep{Jardel2012,Jardel2013a,Jardel2013b,Hagen2019} models of DGs, but can deal also with triaxial systems. 
The weights of the superposed orbits can be used to obtain a numerical estimate of the  DF and thus of the velocity distribution. The galaxy mass distribution can be inferred by comparing the performances of libraries built with different $\Phitot$ \citep{Jardel2012,Jardel2013a,Jardel2013b,Breddels2013a,Breddels2013b,Kowalczyk2017,Kowalczyk2018,Hagen2019,Kowalczyk2019}. These models are flexible, but less intelligible than the analytic-DF models, because the DF is constructed numerically. 

\item {\em Models based on the Jeans equations}. The Jeans equations relate velocity moments of the DF to $\rhostar$ and $\Phitot$, assuming either spherical \citep[][]{Battaglia2008,Strigari2008,Evans2009,Walker2009a,Diakogiannis2017} or axial  (\cite{Hayashi2012,Zhu2016,Hayashi2020}) symmetry.
The method is often applied simply by solving the equation for the second-order velocity moments in spherical symmetry, but the constraining power of Jeans modelling can be improved by considering the equations for higher-order velocity moments \citep{Lokas2002,Lokas2005,Lokas2009,Richardson2013} or by complementing it with the virial shape parameters, related to the projected virial theorem \citep{Richardson2014,Read2018,Read2019,Genina2020a}. 
Jeans-equation methods are versatile, but in general are not guaranteed to provide physical solutions (i.e.\ solutions generated by positive DFs). 

\item {\em Estimators of mass within a given radius}. 
In spherical symmetry, estimates of $\Mdyn$ within a reference radius can be obtained solving the second-order Jeans equation with some simplifying assumptions \citep{Walker2009a,Lazar2020a} or considering the projected virial theorem \citep{Amorisco2012}. These estimates are remarkably robust when the reference radius is $\approx 1.8\Rhalf$ \citep{Wolf2010,Errani2018}. When  distinct stellar components are present, with different $\Rhalf$,  mass estimates at different radii can be used to constrain $\Mdyn(r)$ \citep{Walker2011}. 
These approximated estimators  do not require spatially resolved kinematics, but just measurements of the global l.o.s.\ or plane-of-the-sky stellar velocity dispersion and of $\Rhalf$, and thus are especially useful for UFDs and distant LG dwarfs. 

\end{itemize}

\end{mdframed}

% {\bf End-Box-2}
\bigskip

\section{Inferred dark matter content and distribution}
\label{sec:results}

Within the standard Newtonian framework, it is now well established  that the  LG  DGs are DM dominated systems, in the sense that the density of  DM is typically higher than that of baryons not only in the galaxy outskirts, but also in the central regions \citep[e.g.][]{Mateo1998,Simon2019}. Here we summarise the main properties of LG dwarfs in terms of  DM content and distribution. 

\subsection{Integrated measurements of dark matter mass} 
\label{sec:integratedmass}

\begin{figure}[t!]
\centering
\includegraphics[width=\textwidth]{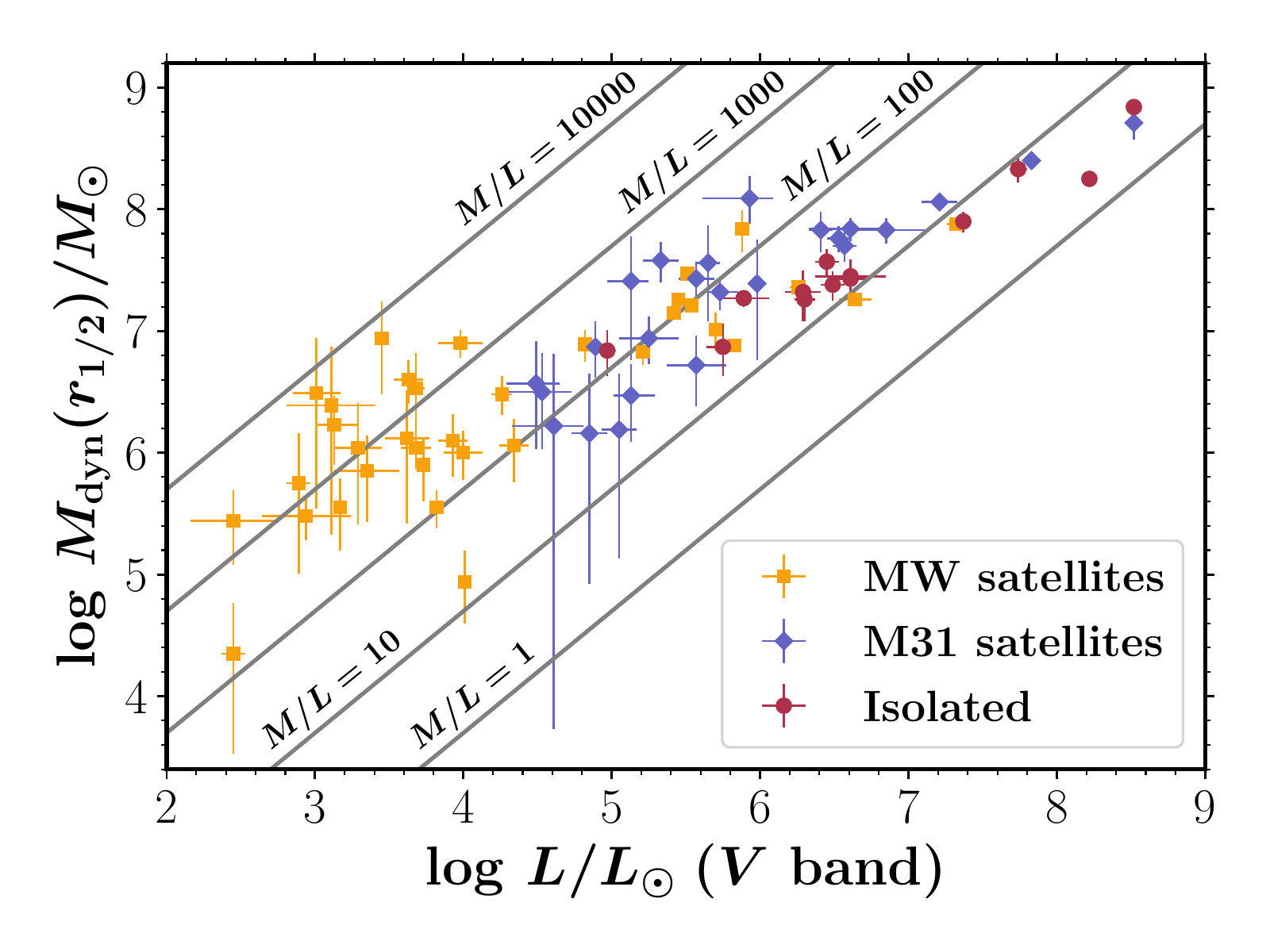}
\caption{Dynamical mass $\Mdyn$ within the 3D half-light radius $\rhalf$ as a function of total $V$-band luminosity $L$ for the LG DGs listed in Table~\ref{tab:mdyn}.
The dynamical masses are estimated using equation~(\ref{eq:massestimator}). The diagonal lines indicate loci of constant mass-to-light ratio $M/L\equiv \Mdyn(\rhalf)/(L/2)$. The values of $M/L$ are in units of $\Msun/\Lsun$ in the $V$ band. 
The symbols represent the estimated median values of $L$ and $\Mdyn$ and the error bars correspond to the 16-th and 84-th percentiles of the distributions of these quantities. The distribution of $\Mdyn$ is obtained from 10000 Monte Carlo realizations in which the values of $\sigmalos$ and $\Rhalf$ are extracted from symmetrized  Gaussian distributions centred on the median observed $\sigmalos$ and $\Rhalf$, and with standard deviation given by the average of the lower (16\%) and upper (84\%) uncertainties in these quantities
(all listed in Table~\ref{tab:mdyn}). 
We neglected the uncertainties in the distance and ellipticity measurements when transforming  the semi-major axis $\ahalf$ of the half-light ellipse in angular units into the circularized half-light radius $\Rhalf$ in physical units.
}
\label{fig:mdyn}
\end{figure}

\begin{table}[th!]
\begin{center}
%\begin{minipage}{174pt}
\begin{minipage}{320pt}
\caption{L.o.s. velocity dispersion $\sigmalos$, total $V$-band luminosity $L$, semi-major axis of the half-light ellipse $a_{1/2}$, ellipticity $\epsilon$, and dynamical mass $\Mdyn(\rhalf)$, estimated in this work using equation~(\ref{eq:massestimator}), of the LG DGs with resolved measurements of $\sigmalos$ shown in Figs.~\ref{fig:sigma} and \ref{fig:mdyn}. 
Uncertainties correspond to 16\%-84\% confidence intervals.   
}
\label{tab:mdyn}
\begin{tabular}{@{}lllllll@{}}
\toprule
Name & $\sigmalos$ & $\log L$  & a$_{1/2}$ & $\epsilon$ & $\log\Mdyn(\rhalf)$ & References \\
  & [km s$^{-1}$] & [$\Lsun$] & [pc] & & [$\Msun$] & \\
\midrule
{\it MW satellites} & & & & & &  \\
            AntliaII & $ 5.7^{+ 1.1}_{- 1.1}$ & $5.88^{+0.03}_{-0.03}$ & $2920^{+ 280}_{- 280}$ &  0.4 & $7.84^{+0.15}_{-0.19}$ &  \citep{Battaglia2021}, \citep{Ji2021}, \citep{Battaglia2021}, \citep{Battaglia2021} \\
          AquariusII & $ 5.4^{+ 3.4}_{- 0.9}$ & $3.68^{+0.06}_{-0.06}$ & $ 159^{+  25}_{-  25}$ &  0.4 & $6.53^{+0.29}_{-0.44}$ &  \citep{Battaglia2021}, \citep{Torrealba2016Aqu2}, \citep{Battaglia2021}, \citep{Battaglia2021} \\
             BootesI & $ 2.4^{+ 0.9}_{- 0.5}$ & $4.34^{+0.10}_{-0.10}$ & $ 247^{+  13}_{-  13}$ &  0.3 & $6.06^{+0.22}_{-0.30}$ &  \citep{Battaglia2021}, \citep{Munoz2018}, \citep{Battaglia2021}, \citep{Battaglia2021} \\
            BootesII & $10.5^{+ 7.4}_{- 7.4}$ & $3.11^{+0.30}_{-0.30}$ & $  31^{+   9}_{-   9}$ &  0.3 & $6.39^{+0.48}_{-1.06}$ &  \citep{Battaglia2021}, \citep{Munoz2018}, \citep{Battaglia2021}, \citep{Battaglia2021} \\
      CanesVenaticiI & $ 7.6^{+ 0.4}_{- 0.4}$ & $5.45^{+0.02}_{-0.02}$ & $ 458^{+  12}_{-  12}$ &  0.4 & $7.26^{+0.05}_{-0.05}$ &  \citep{Battaglia2021}, \citep{Munoz2018}, \citep{Battaglia2021}, \citep{Battaglia2021} \\
     CanesVenaticiII & $ 4.6^{+ 1.0}_{- 1.0}$ & $4.00^{+0.13}_{-0.13}$ & $  66^{+  11}_{-  11}$ &  0.4 & $6.00^{+0.18}_{-0.22}$ &  \citep{Battaglia2021}, \citep{Munoz2018}, \citep{Battaglia2021}, \citep{Battaglia2021} \\
              Carina & $ 6.6^{+ 1.2}_{- 1.2}$ & $5.70^{+0.02}_{-0.02}$ & $ 313^{+   3}_{-   3}$ &  0.4 & $7.01^{+0.14}_{-0.17}$ &  \citep{Battaglia2021}, \citep{Munoz2018}, \citep{Battaglia2021}, \citep{Battaglia2021} \\
            CarinaII & $ 3.4^{+ 1.2}_{- 0.8}$ & $3.73^{+0.04}_{-0.04}$ & $  91^{+   7}_{-   7}$ &  0.3 & $5.90^{+0.22}_{-0.30}$ & \citep{Battaglia2021}, \citep{Torrealba2018}, \citep{Battaglia2021}, \citep{Battaglia2021}  \\
           CarinaIII & $ 5.6^{+ 4.3}_{- 2.1}$ & $2.89^{+0.08}_{-0.08}$ & $  30^{+   8}_{-   8}$ &  0.6 & $5.75^{+0.41}_{-0.74}$ &  \citep{Battaglia2021}, \citep{Torrealba2018}, \citep{Battaglia2021}, \citep{Battaglia2021} \\
       ComaBerenices & $ 4.6^{+ 0.8}_{- 0.8}$ & $3.68^{+0.10}_{-0.10}$ & $  69^{+   3}_{-   3}$ &  0.4 & $6.04^{+0.14}_{-0.17}$ &  \citep{Battaglia2021}, \citep{Munoz2018}, \citep{Battaglia2021}, \citep{Battaglia2021} \\
            CraterII & $ 2.7^{+ 0.3}_{- 0.3}$ & $5.21^{+0.04}_{-0.04}$ & $1057^{+  84}_{-  84}$ &  0.1 & $6.83^{+0.10}_{-0.11}$ & \citep{Battaglia2021}, \citep{Torrealba2016}, \citep{Battaglia2021}, \citep{Battaglia2021}  \\
               Draco & $ 9.0^{+ 0.3}_{- 0.3}$ & $5.42^{+0.02}_{-0.02}$ & $ 225^{+   2}_{-   2}$ &  0.3 & $7.15^{+0.03}_{-0.03}$ &  \citep{Battaglia2021}, \citep{Munoz2018}, \citep{Battaglia2021}, \citep{Battaglia2021} \\
          EridanusII & $ 6.9^{+ 1.2}_{- 0.9}$ & $4.82^{+0.04}_{-0.04}$ & $ 243^{+  12}_{-  12}$ &  0.5 & $6.89^{+0.12}_{-0.14}$ &  \citep{Battaglia2021}, \citep{Munoz2018}, \citep{Battaglia2021}, \citep{Battaglia2021} \\
              Fornax & $11.4^{+ 0.4}_{- 0.4}$ & $7.32^{+0.06}_{-0.06}$ & $ 749^{+   4}_{-   4}$ &  0.3 & $7.88^{+0.03}_{-0.03}$ &  \citep{Battaglia2021}, \citep{Munoz2018}, \citep{Battaglia2021}, \citep{Battaglia2021} \\
            Hercules & $ 5.1^{+ 0.9}_{- 0.9}$ & $4.26^{+0.07}_{-0.07}$ & $ 231^{+  25}_{-  25}$ &  0.7 & $6.48^{+0.15}_{-0.17}$ & \citep{Battaglia2021}, \citep{Munoz2018}, \citep{Battaglia2021}, \citep{Battaglia2021}  \\
         HorologiumI & $ 4.9^{+ 2.8}_{- 0.9}$ & $3.35^{+0.22}_{-0.22}$ & $  39^{+   8}_{-   8}$ &  0.3 & $5.85^{+0.29}_{-0.42}$ &  \citep{Battaglia2021}, \citep{Munoz2018}, \citep{Battaglia2021}, \citep{Battaglia2021} \\
             HydrusI & $ 2.7^{+ 0.5}_{- 0.4}$ & $3.82^{+0.03}_{-0.03}$ & $  59^{+   4}_{-   4}$ &  0.2 & $5.55^{+0.14}_{-0.17}$ & \citep{Battaglia2021}, \citep{Koposov2018}, \citep{Battaglia2021}, \citep{Battaglia2021}  \\
                LeoI & $ 9.2^{+ 0.4}_{- 0.4}$ & $6.64^{+0.11}_{-0.11}$ & $ 276^{+   2}_{-   2}$ &  0.3 & $7.26^{+0.04}_{-0.04}$ & \citep{Battaglia2021}, \citep{Munoz2018}, \citep{Battaglia2021}, \citep{Battaglia2021}  \\
               LeoII & $ 7.4^{+ 0.4}_{- 0.4}$ & $5.83^{+0.02}_{-0.02}$ & $ 155^{+   1}_{-   1}$ &  0.1 & $6.88^{+0.04}_{-0.05}$ &  \citep{Battaglia2021}, \citep{Munoz2018}, \citep{Battaglia2021}, \citep{Battaglia2021} \\
               LeoIV & $ 3.6^{+ 1.0}_{- 1.1}$ & $3.93^{+0.10}_{-0.10}$ & $ 117^{+  14}_{-  14}$ &  0.2 & $6.10^{+0.22}_{-0.30}$ &  \citep{Battaglia2021}, \citep{Munoz2018}, \citep{Battaglia2021}, \citep{Battaglia2021} \\
          PegasusIII & $ 5.4^{+ 3.0}_{- 2.5}$ & $3.29^{+0.16}_{-0.16}$ & $  53^{+  13}_{-  13}$ &  0.4 & $6.04^{+0.37}_{-0.63}$ &  \citep{Battaglia2021}, \citep{Kim2016}, \citep{Battaglia2021}, \citep{Battaglia2021} \\
           PhoenixII & $11.0^{+ 9.4}_{- 5.3}$ & $3.01^{+0.16}_{-0.16}$ & $  36^{+   7}_{-   7}$ &  0.4 & $6.49^{+0.45}_{-0.95}$ &  \citep{Battaglia2021}, \citep{Mutlu-Pakdil2018}, \citep{Battaglia2021}, \citep{Battaglia2021} \\
            PiscesII & $ 5.4^{+ 3.6}_{- 2.4}$ & $3.62^{+0.15}_{-0.15}$ & $  62^{+  10}_{-  10}$ &  0.4 & $6.12^{+0.39}_{-0.70}$ & \citep{Battaglia2021}, \citep{Munoz2018}, \citep{Battaglia2021}, \citep{Battaglia2021}  \\
         ReticulumII & $ 3.2^{+ 1.6}_{- 0.5}$ & $3.17^{+0.04}_{-0.04}$ & $  57^{+   3}_{-   3}$ &  0.6 & $5.55^{+0.24}_{-0.35}$ & \citep{Battaglia2021}, \citep{Mutlu-Pakdil2018}, \citep{Battaglia2021}, \citep{Battaglia2021}  \\
            Sculptor & $10.1^{+ 0.3}_{- 0.3}$ & $6.26^{+0.06}_{-0.06}$ & $ 303^{+   4}_{-   4}$ &  0.4 & $7.36^{+0.03}_{-0.03}$ & \citep{Battaglia2021}, \citep{Munoz2018}, \citep{Battaglia2021}, \citep{Battaglia2021}  \\
              Segue1 & $ 3.7^{+ 1.4}_{- 1.1}$ & $2.45^{+0.29}_{-0.29}$ & $  26^{+   2}_{-   2}$ &  0.3 & $5.44^{+0.25}_{-0.36}$ &  \citep{Battaglia2021}, \citep{Munoz2018}, \citep{Battaglia2021}, \citep{Battaglia2021} \\
             Sextans & $ 8.4^{+ 0.4}_{- 0.4}$ & $5.51^{+0.04}_{-0.04}$ & $ 527^{+  17}_{-  14}$ &  0.3 & $7.47^{+0.04}_{-0.04}$ &  \citep{Battaglia2021}, \citep{Cicuendez2018first}, \citep{Battaglia2021}, \citep{Battaglia2021} \\
            TucanaII & $ 8.6^{+ 4.4}_{- 2.7}$ & $3.45^{+0.04}_{-0.04}$ & $ 164^{+  27}_{-  18}$ &  0.4 & $6.94^{+0.30}_{-0.46}$ & \citep{Battaglia2021}, \citep{Koposov2015}, \citep{Battaglia2021}, \citep{Battaglia2021}  \\
           TucanaIII & $ 0.9^{+ 0.6}_{- 0.5}$ & $2.45^{+0.08}_{-0.08}$ & $  33^{+   7}_{-   7}$ &  0.2 & $4.35^{+0.42}_{-0.82}$ &  \citep{Battaglia2021}, \citep{Mutlu-Pakdil2018}, \citep{Battaglia2021}, \citep{Battaglia2021} \\
            TucanaIV & $ 4.3^{+ 1.7}_{- 1.0}$ & $3.13^{+0.16}_{-0.12}$ & $ 127^{+  19}_{-  12}$ &  0.4 & $6.23^{+0.24}_{-0.33}$ &  \citep{Battaglia2021}, \citep{Simon2020}, \citep{Battaglia2021}, \citep{Battaglia2021} \\
          UrsaMajorI & $ 7.6^{+ 1.0}_{- 1.0}$ & $3.98^{+0.15}_{-0.15}$ & $ 230^{+   8}_{-   8}$ &  0.6 & $6.90^{+0.11}_{-0.12}$ &  \citep{Battaglia2021}, \citep{Munoz2018}, \citep{Battaglia2021}, \citep{Battaglia2021} \\
         UrsaMajorII & $ 6.7^{+ 1.4}_{- 1.4}$ & $3.63^{+0.10}_{-0.10}$ & $ 140^{+   4}_{-   4}$ &  0.6 & $6.60^{+0.16}_{-0.20}$ & \citep{Battaglia2021}, \citep{Munoz2018}, \citep{Battaglia2021}, \citep{Battaglia2021}  \\
           UrsaMinor & $ 8.0^{+ 0.3}_{- 0.3}$ & $5.54^{+0.02}_{-0.02}$ & $ 403^{+   2}_{-   2}$ &  0.6 & $7.21^{+0.03}_{-0.03}$ & \citep{Battaglia2021}, \citep{Munoz2018}, \citep{Battaglia2021}, \citep{Battaglia2021}  \\
            Willman1 & $ 4.0^{+ 0.8}_{- 0.8}$ & $2.94^{+0.30}_{-0.30}$ & $  27^{+   2}_{-   2}$ &  0.5 & $5.48^{+0.16}_{-0.20}$ & \citep{Battaglia2021}, \citep{Munoz2018}, \citep{Battaglia2021}, \citep{Battaglia2021}  \\
\botrule
\end{tabular}
\end{minipage}
\end{center}
\end{table}
\setcounter{table}{0}
\begin{table}[th!]
\begin{center}
%\begin{minipage}{174pt}
\begin{minipage}{320pt}
\caption{(Continued)}
%\label{tab1}
\begin{tabular}{@{}lllllll@{}}
%\begin{tabular}{@{}llll@{}}
\toprule
Name & $\sigmalos$ & $\log L$  & a$_{1/2}$ & $\epsilon$ & $\log\Mdyn(\rhalf)$ & References \\
  & [km s$^{-1}$] & [$\Lsun$] & [pc] & & [$\Msun$] & \\
\midrule
{\it Isolated DGs} & & & & & & \\
            AndXVIII & $ 9.7^{+ 2.3}_{- 2.3}$ & $6.29^{+0.12}_{-0.12}$ & $ 316^{+  22}_{-  22}$ &  0.4 & $7.32^{+0.18}_{-0.24}$ & \citep{Higgs2021}, \citep{Higgs2021}, \citep{Higgs2021}, \citep{Higgs2021}  \\
            Aquarius & $ 7.8^{+ 1.8}_{- 1.1}$ & $6.30^{+0.07}_{-0.07}$ & $ 468^{+  26}_{-  26}$ &  0.5 & $7.26^{+0.15}_{-0.18}$ & \citep{Higgs2021}, \citep{Higgs2021}, \citep{Higgs2021}, \citep{Higgs2021}  \\
               Cetus & $ 8.3^{+ 1.0}_{- 1.0}$ & $6.45^{+0.08}_{-0.08}$ & $ 703^{+  32}_{-  32}$ &  0.3 & $7.57^{+0.10}_{-0.11}$ &  \citep{Higgs2021}, \citep{Higgs2021}, \citep{Higgs2021}, \citep{Higgs2021} \\
              IC1613 & $11.3^{+ 0.5}_{- 0.5}$ & $8.22^{+0.04}_{-0.04}$ & $1670^{+  11}_{-  11}$ &  0.2 & $8.25^{+0.04}_{-0.04}$ &  \citep{Battaglia2021}, \citep{Higgs2021}, \citep{Battaglia2021}, \citep{Battaglia2021} \\               
                LeoA & $ 9.0^{+ 0.8}_{- 0.6}$ & $6.61^{+0.12}_{-0.12}$ & $ 480^{+ 190}_{- 190}$ &  0.4 & $7.43^{+0.16}_{-0.22}$ &  \citep{Battaglia2021}, \citep{Higgs2021}, \citep{Battaglia2021}, \citep{Battaglia2021} \\
                LeoT & $ 7.5^{+ 1.6}_{- 1.6}$ & $4.97^{+0.06}_{-0.06}$ & $ 151^{+  15}_{-  15}$ &  0.2 & $6.84^{+0.17}_{-0.21}$ &  \citep{Battaglia2021}, \citep{Munoz2018}, \citep{Battaglia2021}, \citep{Battaglia2021} \\
             NGC6822 & $23.2^{+ 1.2}_{- 1.2}$ & $8.52^{+0.03}_{-0.03}$ & $1633^{+  95}_{-  95}$ &  0.3 & $8.84^{+0.05}_{-0.05}$ &  \citep{Battaglia2021}, \citep{Higgs2021}, \citep{Battaglia2021}, \citep{Battaglia2021} \\
            Peg-dIrr & $12.3^{+ 1.2}_{- 1.1}$ & $7.37^{+0.04}_{-0.04}$ & $ 840^{+  11}_{-  11}$ &  0.6 & $7.90^{+0.08}_{-0.09}$ &  \citep{Battaglia2021}, \citep{Higgs2021}, \citep{Battaglia2021}, \citep{Battaglia2021} \\
            Phoenix & $ 9.3^{+ 0.7}_{- 0.7}$ & $5.89^{+0.17}_{-0.17}$ & $ 273^{+   8}_{-   8}$ &  0.3 & $7.27^{+0.06}_{-0.07}$ & \citep{Battaglia2021}, \citep{Higgs2021}, \citep{Battaglia2021}, \citep{Battaglia2021}  \\
            Sg-dIrr & $ 9.4^{+ 1.5}_{- 1.1}$ & $6.49^{+0.08}_{-0.08}$ & $ 443^{+  24}_{-  24}$ &  0.6 & $7.38^{+0.11}_{-0.13}$ &  \citep{Battaglia2021}, \citep{Higgs2021}, \citep{Battaglia2021}, \citep{Battaglia2021} \\
            Tucana & $ 6.2^{+ 1.6}_{- 1.3}$ & $5.75^{+0.11}_{-0.11}$ & $ 286^{+  36}_{-  36}$ &  0.5 & $6.87^{+0.19}_{-0.24}$ &  \citep{Taibi2020}, \citep{Saviane1996}, \citep{Taibi2020}, \citep{Taibi2020} \\
            UGC4879 & $ 9.6^{+ 1.3}_{- 1.2}$ & $6.61^{+0.24}_{-0.24}$ & $ 435^{+  38}_{-  38}$ &  0.4 & $7.45^{+0.11}_{-0.13}$ &  \citep{Battaglia2021}, \citep{Higgs2021}, \citep{Battaglia2021}, \citep{Battaglia2021} \\
            WLM & $17.5^{+ 2.0}_{- 2.0}$ & $7.74^{+0.06}_{-0.06}$ & $1113^{+  35}_{-  35}$ &  0.5 & $8.33^{+0.09}_{-0.11}$ &  \citep{Battaglia2021}, \citep{Higgs2021}, \citep{Battaglia2021}, \citep{Battaglia2021} \\

             \hline
{\it M31 satellites} & & & & & & \\
                AndI & $10.2^{+ 1.9}_{- 1.9}$ & $6.41^{+0.08}_{-0.08}$ & $ 815^{+  40}_{-  40}$ &  0.3 & $7.83^{+0.15}_{-0.18}$ & \citep{Higgs2021}, \citep{Higgs2021}, \citep{Higgs2021}, \citep{Higgs2021}  \\ 
               AndII & $ 7.8^{+ 1.1}_{- 1.1}$ & $6.57^{+0.08}_{-0.08}$ & $ 965^{+  45}_{-  45}$ &  0.2 & $7.70^{+0.11}_{-0.13}$ & \citep{Higgs2021}, \citep{Higgs2021}, \citep{Higgs2021}, \citep{Higgs2021}  \\   
              AndIII & $ 9.3^{+ 1.4}_{- 1.4}$ & $5.73^{+0.12}_{-0.12}$ & $ 405^{+  35}_{-  35}$ &  0.6 & $7.32^{+0.12}_{-0.15}$ &  \citep{Higgs2021}, \citep{Higgs2021}, \citep{Higgs2021}, \citep{Higgs2021} \\   
                AndV & $11.5^{+ 5.4}_{- 4.4}$ & $5.65^{+0.08}_{-0.08}$ & $ 345^{+  40}_{-  40}$ &  0.3 & $7.56^{+0.31}_{-0.48}$ & \citep{Higgs2021}, \citep{Higgs2021}, \citep{Higgs2021}, \citep{Higgs2021}  \\    
               AndVI & $12.4^{+ 1.5}_{- 1.3}$ & $6.53^{+0.08}_{-0.08}$ & $ 524^{+  49}_{-  49}$ &  0.4 & $7.76^{+0.10}_{-0.11}$ &  \citep{Higgs2021}, \citep{Higgs2021}, \citep{Higgs2021}, \citep{Higgs2021}  \\ 
              AndVII & $13.0^{+ 1.0}_{- 1.0}$ & $7.21^{+0.12}_{-0.12}$ & $ 775^{+  43}_{-  43}$ &  0.1 & $8.06^{+0.07}_{-0.07}$ & \citep{Higgs2021}, \citep{Higgs2021}, \citep{Higgs2021}, \citep{Higgs2021}  \\   
               AndIX & $10.9^{+ 2.0}_{- 2.0}$ & $5.33^{+0.12}_{-0.12}$ & $ 348^{+  52}_{-  34}$ &  0.0 & $7.58^{+0.15}_{-0.18}$ & \citep{Tollerud2012}, \citep{Martin2016}, \citep{Martin2016}, \citep{Martin2016}  \\
                AndX & $ 6.4^{+ 1.4}_{- 1.4}$ & $4.89^{+0.12}_{-0.12}$ & $ 214^{+  77}_{-  38}$ &  0.1 & $6.87^{+0.21}_{-0.25}$ &  \citep{Tollerud2012}, \citep{Martin2016}, \citep{Martin2016}, \citep{Martin2016} \\   
             AndXIII & $ 5.8^{+ 2.0}_{- 2.0}$ & $4.53^{+0.20}_{-0.28}$ & $ 176^{+  88}_{-  66}$ &  0.6 & $6.50^{+0.32}_{-0.47}$ & \citep{Tollerud2012}, \citep{Martin2016}, \citep{Martin2016}, \citep{Martin2016}  \\      
               AndXV & $ 4.0^{+ 1.4}_{- 1.4}$ & $5.13^{+0.16}_{-0.12}$ & $ 230^{+  35}_{-  25}$ &  0.2 & $6.47^{+0.26}_{-0.38}$ &  \citep{Higgs2021}, \citep{Higgs2021}, \citep{Higgs2021}, \citep{Higgs2021} \\           
              AndXVI & $ 3.8^{+ 2.9}_{- 2.9}$ & $4.85^{+0.12}_{-0.12}$ & $ 130^{+  30}_{-  15}$ &  0.3 & $6.16^{+0.49}_{-1.24}$ & \citep{Higgs2021}, \citep{Higgs2021}, \citep{Higgs2021}, \citep{Higgs2021}   \\  
             AndXVII & $ 2.9^{+ 2.2}_{- 1.9}$ & $5.05^{+0.12}_{-0.12}$ & $ 285^{+  55}_{-  45}$ &  0.5 & $6.19^{+0.46}_{-1.06}$ & \citep{Higgs2021}, \citep{Higgs2021}, \citep{Higgs2021}, \citep{Higgs2021}  \\
              AndXIX & $ 7.8^{+ 1.7}_{- 1.5}$ & $5.93^{+0.16}_{-0.32}$ & $3390^{+ 810}_{- 450}$ &  0.6 & $8.09^{+0.18}_{-0.21}$ &  \citep{Collins2020}, \citep{Martin2016}, \citep{Martin2016}, \citep{Martin2016} \\   
               AndXX & $ 7.1^{+ 3.9}_{- 2.5}$ & $4.49^{+0.16}_{-0.20}$ & $  90^{+  35}_{-  20}$ &  0.1 & $6.57^{+0.35}_{-0.54}$ & \citep{Higgs2021}, \citep{Higgs2021}, \citep{Higgs2021}, \citep{Higgs2021}  \\              
              AndXXI & $ 6.1^{+ 1.0}_{- 0.9}$ & $5.57^{+0.12}_{-0.12}$ & $ 987^{+ 192}_{-  96}$ &  0.4 & $7.43^{+0.14}_{-0.16}$ & \citep{Collins2021}, \citep{Martin2016}, \citep{Martin2016}, \citep{Martin2016}  \\
             AndXXII & $ 3.5^{+ 4.2}_{- 2.5}$ & $4.61^{+0.20}_{-0.28}$ & $ 240^{+  80}_{-  53}$ &  0.6 & $6.22^{+0.59}_{-2.49}$ & \citep{Tollerud2012}, \citep{Martin2016}, \citep{Martin2016}, \citep{Martin2016}   \\
           AndXXVIII & $ 4.9^{+ 1.6}_{- 1.6}$ & $5.57^{+0.20}_{-0.20}$ & $ 310^{+  13}_{-  13}$ &  0.4 & $6.72^{+0.24}_{-0.34}$ & \citep{Tollerud2013}, \citep{Higgs2021}, \citep{Higgs2021SOLO2}, \citep{Higgs2021SOLO2}  \\
             AndXXIX & $ 5.7^{+ 1.2}_{- 1.2}$ & $5.25^{+0.20}_{-0.20}$ & $ 362^{+  57}_{-  55}$ &  0.3 & $6.94^{+0.18}_{-0.21}$ & \citep{Higgs2021}, \citep{Higgs2021}, \citep{Higgs2021}, \citep{Higgs2021}  \\   
              AndXXX & $11.8^{+ 7.7}_{- 4.7}$ & $5.13^{+0.12}_{-0.16}$ & $ 270^{+  50}_{-  50}$ &  0.4 & $7.41^{+0.37}_{-0.65}$ & \citep{Higgs2021}, \citep{Higgs2021}, \citep{Higgs2021}, \citep{Higgs2021}  \\    
             AndXXXI & $10.3^{+ 0.9}_{- 0.9}$ & $6.61^{+0.28}_{-0.28}$ & $ 930^{+ 100}_{- 120}$ &  0.4 & $7.84^{+0.09}_{-0.10}$ & \citep{Higgs2021}, \citep{Higgs2021}, \citep{Higgs2021}, \citep{Higgs2021}   \\              
            AndXXXII & $ 8.4^{+ 0.6}_{- 0.6}$ & $6.85^{+0.28}_{-0.28}$ & $1470^{+ 290}_{- 240}$ &  0.5 & $7.83^{+0.10}_{-0.11}$ & \citep{Higgs2021}, \citep{Higgs2021}, \citep{Higgs2021}, \citep{Higgs2021}  \\
            LGS3 & $ 7.9^{+ 5.3}_{- 2.9}$ & $5.98^{+0.05}_{-0.05}$ & $ 469^{+  44}_{-  44}$ &  0.2 & $7.39^{+0.36}_{-0.63}$ &  \citep{Cook1999}, \citep{Higgs2021}, \citep{McConnachie2012}, \citep{Lee1995}  \\
              NGC185 & $24.0^{+ 1.0}_{- 1.0}$ & $7.83^{+0.05}_{-0.05}$ & $ 529^{+   7}_{-   7}$ &  0.2 & $8.40^{+0.04}_{-0.04}$ &  \citep{Geha2010}, \citep{Higgs2021}, \citep{Crnojevic2014}, \citep{Crnojevic2014} \\
              NGC205 & $35.0^{+ 5.0}_{- 5.0}$ & $8.52^{+0.05}_{-0.05}$ & $ 589^{+  23}_{-  23}$ &  0.4 & $8.71^{+0.11}_{-0.14}$ &   \citep{Geha2006}, \citep{Higgs2021}, \citep{McConnachie2012}, \citep{McConnachie2012} \\                              
\botrule
\end{tabular}
\end{minipage}
\end{center}
\end{table}

\begin{table}[th!]
\begin{center}
\caption{Upper limit  on l.o.s.\ velocity dispersion $\sigmalosmax$, total $V$-band luminosity $L$ and semi-major axis of the half-light ellipse  $a_{1/2}$ for the systems with unresolved measurements of $\sigmalos$. These systems, which are shown in Fig.~\ref{fig:sigma} but not in Fig.~\ref{fig:mdyn}, are all MW satellites. The column C.I. gives the value of the confidence interval to which the upper limit on $\sigmalos$ refers to. 
Uncertainties correspond to 16\%-84\% confidence intervals, except when noted otherwise.   
}
\label{tab:unresolved}
\begin{tabular}{@{}llllll@{}}
\hline\hline
Name & $\sigmalosmax$ & $\log L$  & a$_{1/2}$ &  C.I. & References\\
  & [km s$^{-1}$] & [$\Lsun$] & [pc] & &  \\
  \hline
            ColumbaI &  16.1 & $3.73^{+0.07}_{-0.07}$ & $ 116^{+  10}_{-  10}$ & 90\%  & \cite{Battaglia2021, Drlica-Wagner2015, Battaglia2021} \\
             DracoII &   5.9 & $2.25^{+0.40}_{-0.16}$ & $  18^{+   4}_{-   3}$ & 95\%  & \cite{Battaglia2021, Longeard2018, Battaglia2021} \\
               GrusI &   9.8 & $3.32^{+0.12}_{-0.24}$ & $ 153^{+  19}_{-  27}$ & 68\%   & \cite{Battaglia2021, Munoz2018, Battaglia2021} \\
              GrusII &   2.0 & $3.49^{+0.09}_{-0.09}$ & $  94^{+   7}_{-   7}$ & 95.5\%   & \cite{Battaglia2021, Drlica-Wagner2015, Battaglia2021} \\
        HorologiumII &  54.6 & $2.56^{+0.41}_{-0.41}$ & $  49^{+  13}_{-  13}$ & 90\%   &  \cite{Battaglia2021, Munoz2018, Battaglia2021} \\
             HydraII &   3.6 & $3.77^{+0.15}_{-0.15}$ & $  72^{+  17}_{-  17}$ & 90\%   & \cite{Battaglia2021, Munoz2018, Battaglia2021} \\
        ReticulumIII &  31.2 & $3.25^{+0.12}_{-0.12}$ & $  63^{+  23}_{-  21}$ & 90\%   & \cite{Battaglia2021, Drlica-Wagner2015, Battaglia2021}  \\
              Segue2 &   2.6 & $2.68^{+0.35}_{-0.35}$ & $  36^{+   2}_{-   2}$ & 95\%   & \cite{Battaglia2021, Munoz2018, Battaglia2021} \\
        TriangulumII &   4.2 & $2.41^{+0.16}_{-0.16}$ & $  19^{+   4}_{-   4}$ & 95\%   & \cite{Battaglia2021, Carlin2017, Battaglia2021} \\
             TucanaV &   7.4 & $2.37^{+0.24}_{-0.20}$ & $  33^{+   9}_{-   6}$ & 95.5\%   & \cite{Battaglia2021, Simon2020, Battaglia2021} \\
             \hline
\end{tabular}
\end{center}
\end{table}

The ability to determine the mass distribution of a DG from observations of a sample of stars  depends heavily on the size of the sample and on the quality of the photometric and spectroscopic data. While for luminous dwarfs with of the order of $10^2-10^3$ measured l.o.s.\ velocities, analytic-DF, Schwarzschild or Jeans modelling (see Box 2)
 can be used to constrain the mass density distribution, for LG DGs with a couple of dozen observed  member stars, only estimates of integrated quantities, such as  $\Mdyn$ within the half-light radius, can be obtained. 
To deduce information on the DM mass from knowledge of the dynamical mass one must estimate the mass of the baryonic matter.  
For stellar populations typical of LG DGs, the stellar mass-to-light ratio is $\Mstar/L\lesssim 2$ in solar units, where $L$ is the $V$-band luminosity (e.g.\ \cite{Woo2008,Simon2019}). For most LG DGs the gas mass $\Mgas$ is lower than $\Mstar$, but even LG dIrrs and dTs typically have $\Mgas/L\lesssim 3$ in solar units \citep[][]{Putman2021}, so high ($\gtrsim10$) values of $\Mdyn/L$ must be ascribed to DM. 

For the sample of LG DGs listed in Table~\ref{tab:mdyn} and considered in Fig.~\ref{fig:sigma}, we computed the dynamical mass $\Mdyn(\rhalf)$ within the 3D half-light radius $\rhalf$ using the mass estimator
\begin{equation}
    \Mdyn(\rhalf)=4\frac{\sigmalos^2\Rhalf}{G}
    \label{eq:massestimator}
\end{equation}
\citep[][]{Wolf2010}, where $\sigmalos$ is the global stellar l.o.s.\ velocity dispersion and $\Rhalf$ is the 2D half-light (or effective) radius. When applying the mass estimator (\ref{eq:massestimator}), which is based on spherical models, to galaxies that appear flattened in the plane of the sky, the definition of $\Rhalf$ is not univocal: here, following \cite{Sanders2016a}, we define $\Rhalf$ as the circularized half-light radius $\Rhalf\equiv\ahalf\sqrt{1-\epsilon}$. 
Fig.~\ref{fig:mdyn} plots, for this sample, $\Mdyn(\rhalf)$ as a function of $L$ and, for reference, loci of constant dynamical mass-to-light ratio $M/L\equiv \Mdyn(\rhalf)/(L/2)$. The vast majority of these DGs have $M/L\gtrsim 10$, which implies that most of the mass is in the form of DM even within $\rhalf$.
$\Mdyn/L$ tends to increase for decreasing $L$, ranging from $\Mdyn/L\sim 10$ for the most luminous dwarfs up to $\Mdyn/L \gtrsim 1000$ for the faintest systems, so lower luminosity DGs are more DM dominated. 
Only the three most luminous DGs of the sample (IC~1613, NGC~6822 and NGC~205, all with $L>10^8\Lsun$)  have $M/L$ significantly lower than 10.

To use the observed DGs as probes of cosmological and galaxy formation models (e.g.\ \cite{Bullock2017}), one would be interested to know the virial (i.e.\ total) masses of their host dark halos.
However, while relatively robust estimates are obtained for $\Mdyn(\rhalf)$, the virial mass of DGs is a poorly constrained quantity, because tracers at large radii are lacking and extrapolations are uncertain due to tidal effects \citep[][]{Errani2018}.

An alternative way to look at deviations 
of $\Mdyn$ from the baryonic mass of galaxies, broadly inspired by Milgrom's MOND \citep[][]{Milgrom1983}, is the radial acceleration relation (RAR; \cite{McGaugh2016b}) between the
observed radial acceleration $\gobs$ and the radial acceleration $\gbar$ (at the same radius) expected from the baryonic distribution. For rotation supported galaxies $\gobs\approx\gbar$ at high $\gbar$, while $\gobs/\gbar$ increases gradually for decreasing $\gbar$, which in the context of Newtonian dynamics is interpreted as the presence of DM.  
Gas-poor LG DGs are in the low-$\gbar$ regime: while the dSphs lie on average on the low-$\gbar$ extrapolation of the RAR, UFDs tend to have, within the uncertainties, $\gobs$ independent of $\gbar$ \citep{Lelli2017}, consistent with the finding \citep{Strigari2008} that UDFs and dSphs have, within their central 300 pc, very similar masses.

\subsection{Dark matter density distribution} 

\begin{figure}[t!]
\centering
\includegraphics[width=0.84\textwidth]{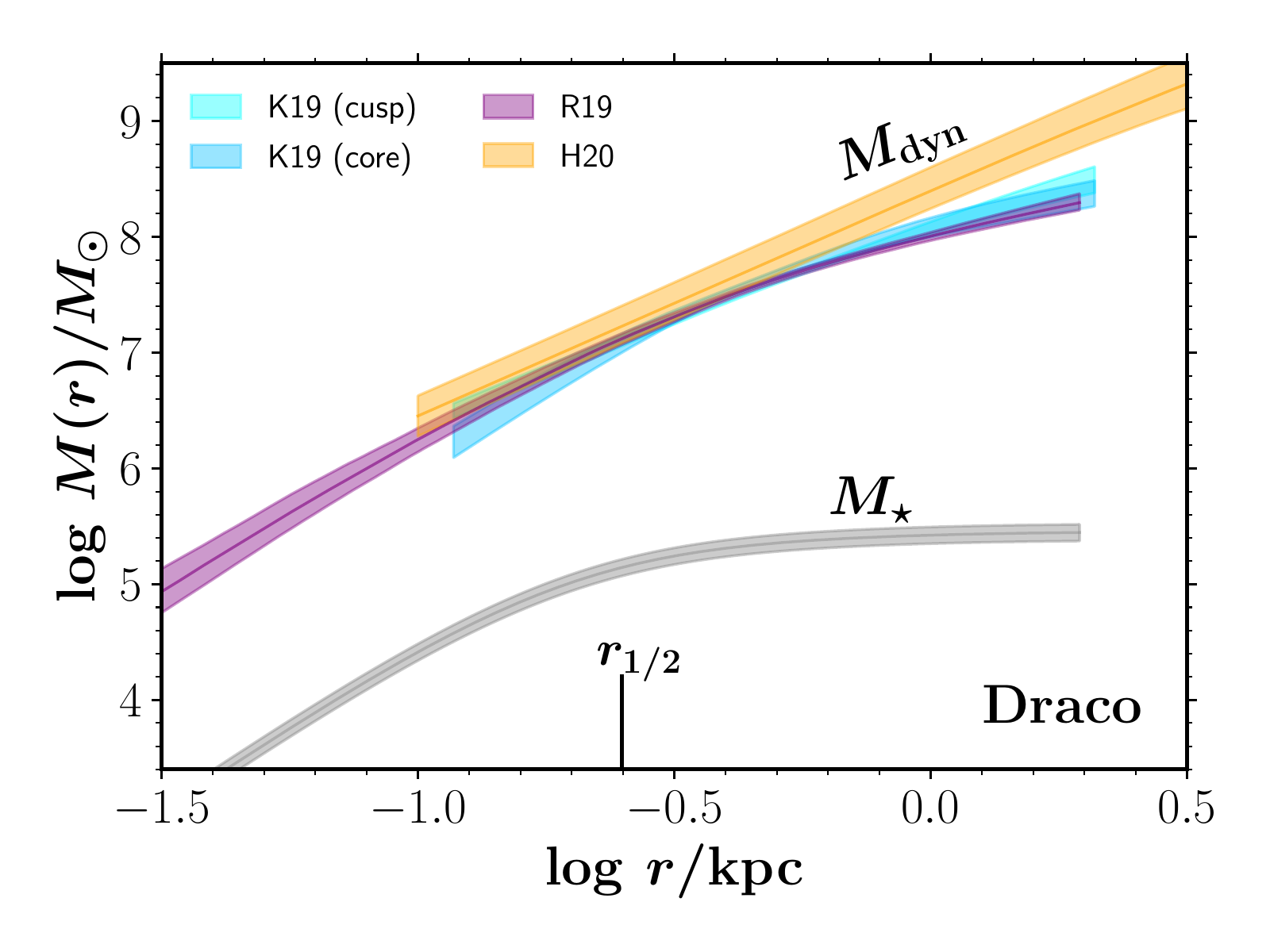}
\includegraphics[width=0.84\textwidth]{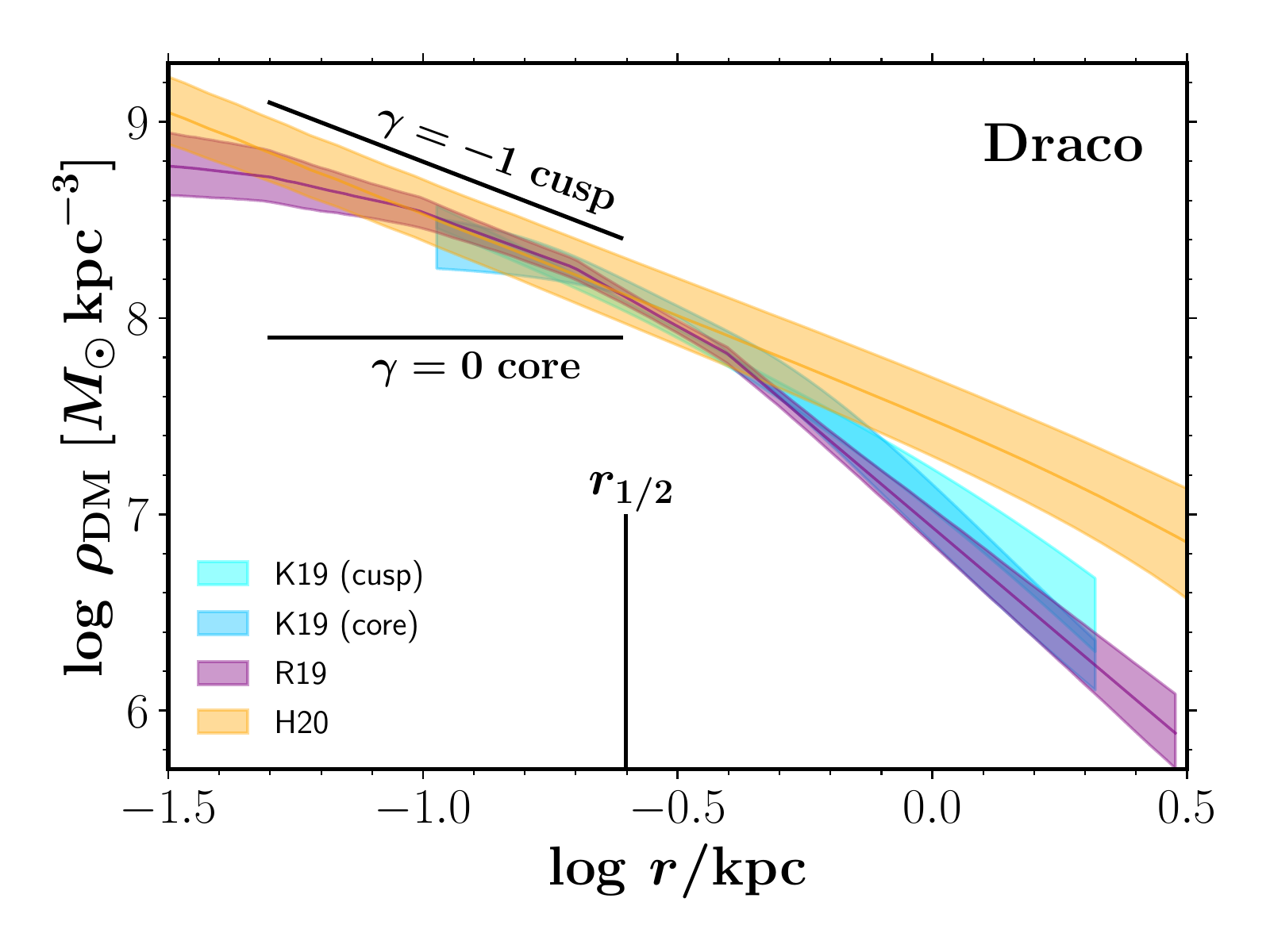}
\caption{
{\em Upper panel.} Mass profiles of the Draco dSph. The coloured bands represent 68\% confidence intervals of the dynamical mass profiles $\Mdyn(r)=\MDM(r)+\Mstar(r)$ as estimated by
Kaplinghat et al.\ (K19, \cite{Kaplinghat2019}; both cuspy and cored models), Read et al.\ (R19, \cite{Read2019}) and Hayashi et al.\ (H20, \cite{Hayashi2020}). In the model of H20 the density distribution has spheroidal symmetry and $r$ is the circularized radius. When available, the median profile is indicated by the solid line. 
The grey solid line and band represent, respectively, the median and 68\% confidence intervals of the stellar mass profile $\Mstar(r)$, as estimated by Read et al. \citep{Read2019}. The vertical black line indicates the 3D stellar half mass radius. {\em Lower panel.} Same as upper panel, but showing the corresponding DM density profiles.  The upper and lower black lines indicate $\rhoDM\propto  r^{\gamma}$ with,  respectively,  $\gamma=-1$ (cusp) and $\gamma=0$ (core) over the radial range $0.2\rhalf<r<\rhalf$.}
\label{fig:draco}
\end{figure}

\begin{figure}[t!]
\centering
\includegraphics[width=0.84\textwidth]{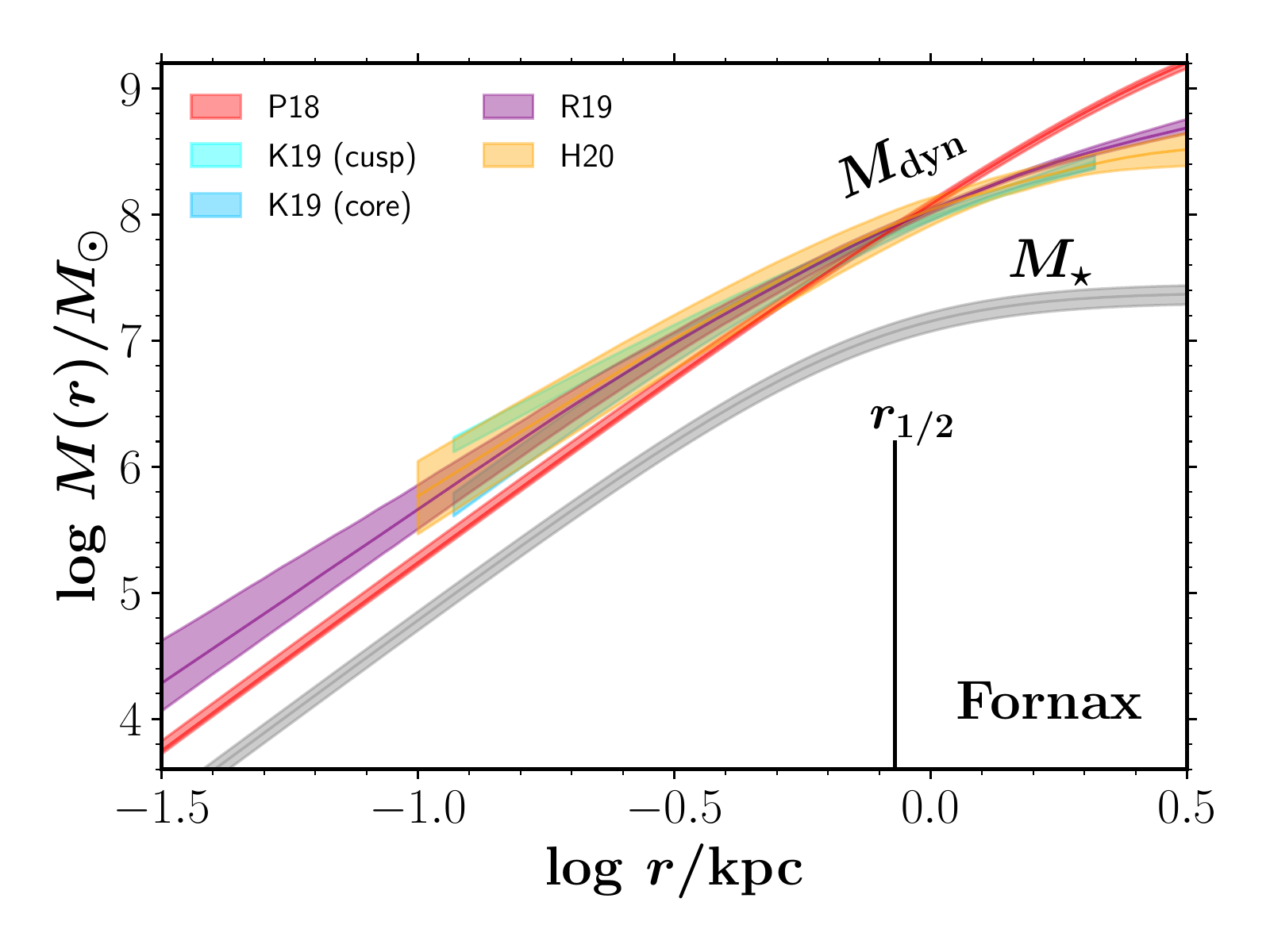}
\includegraphics[width=0.84\textwidth]{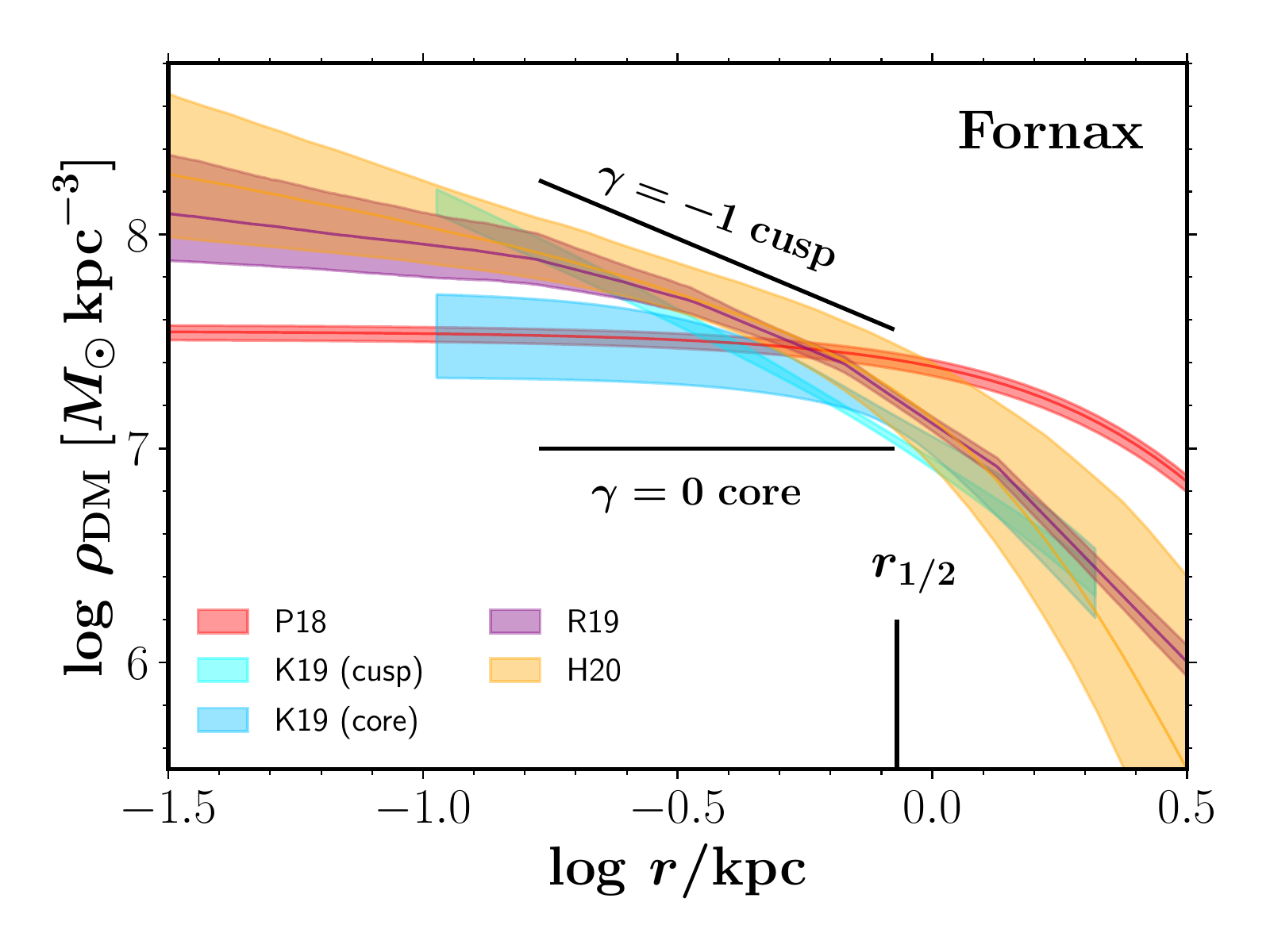}
\caption{
Same as Fig.~\ref{fig:draco}, but for the Fornax dSph. Here the profiles are taken from  Pascale et al.\ (P18, \cite{Pascale2018}), Kaplinghat et al.\  (K19, \cite{Kaplinghat2019}; both cuspy and cored models), Read et al.\ (R19, \cite{Read2019}) and Hayashi et al.\ (H20, \cite{Hayashi2020}; circularized spheroidal model).
}
\label{fig:fornax}
\end{figure}

\begin{figure}[t!]
\centering
\includegraphics[width=0.84\textwidth]{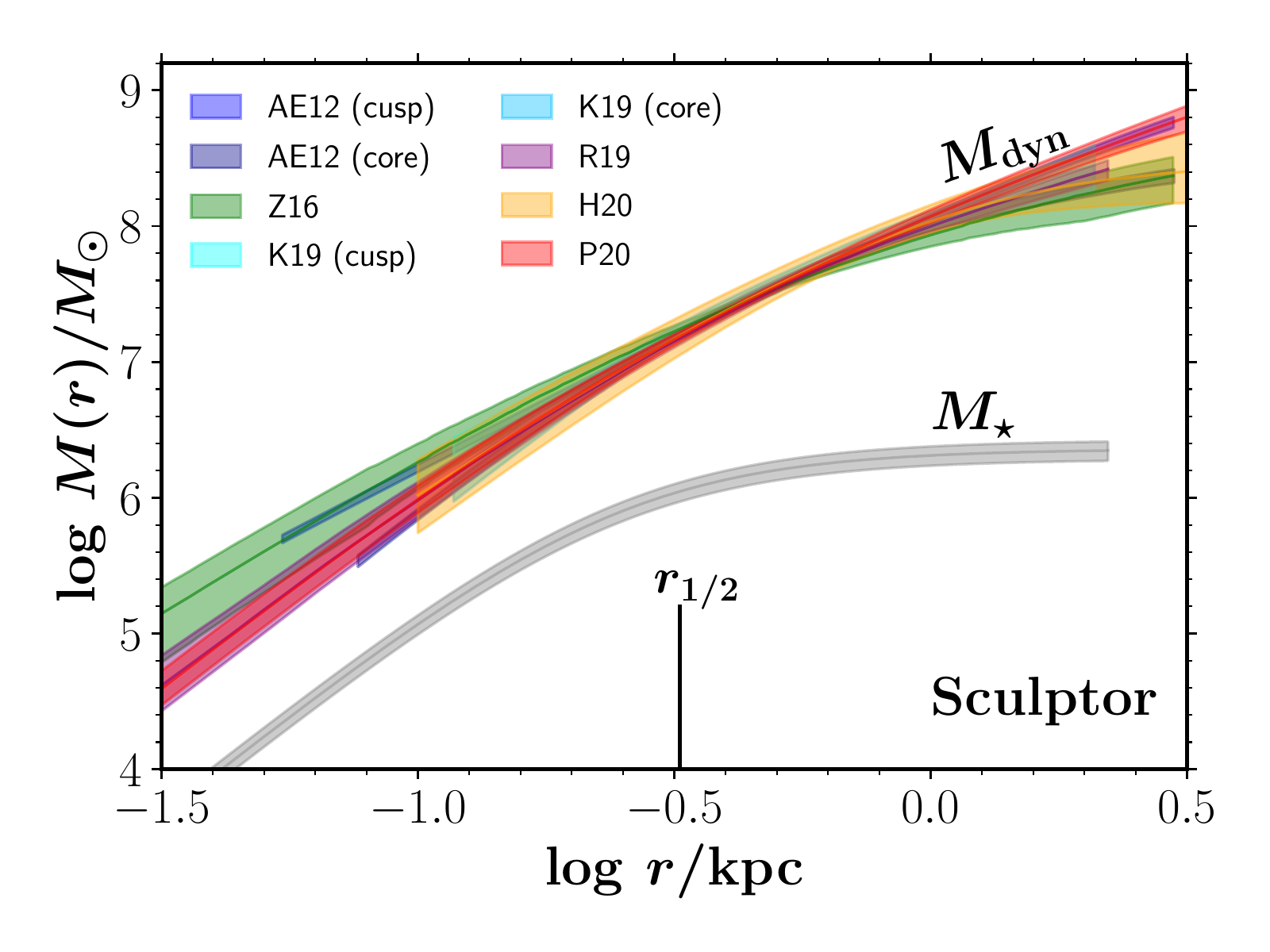}
\includegraphics[width=0.84\textwidth]{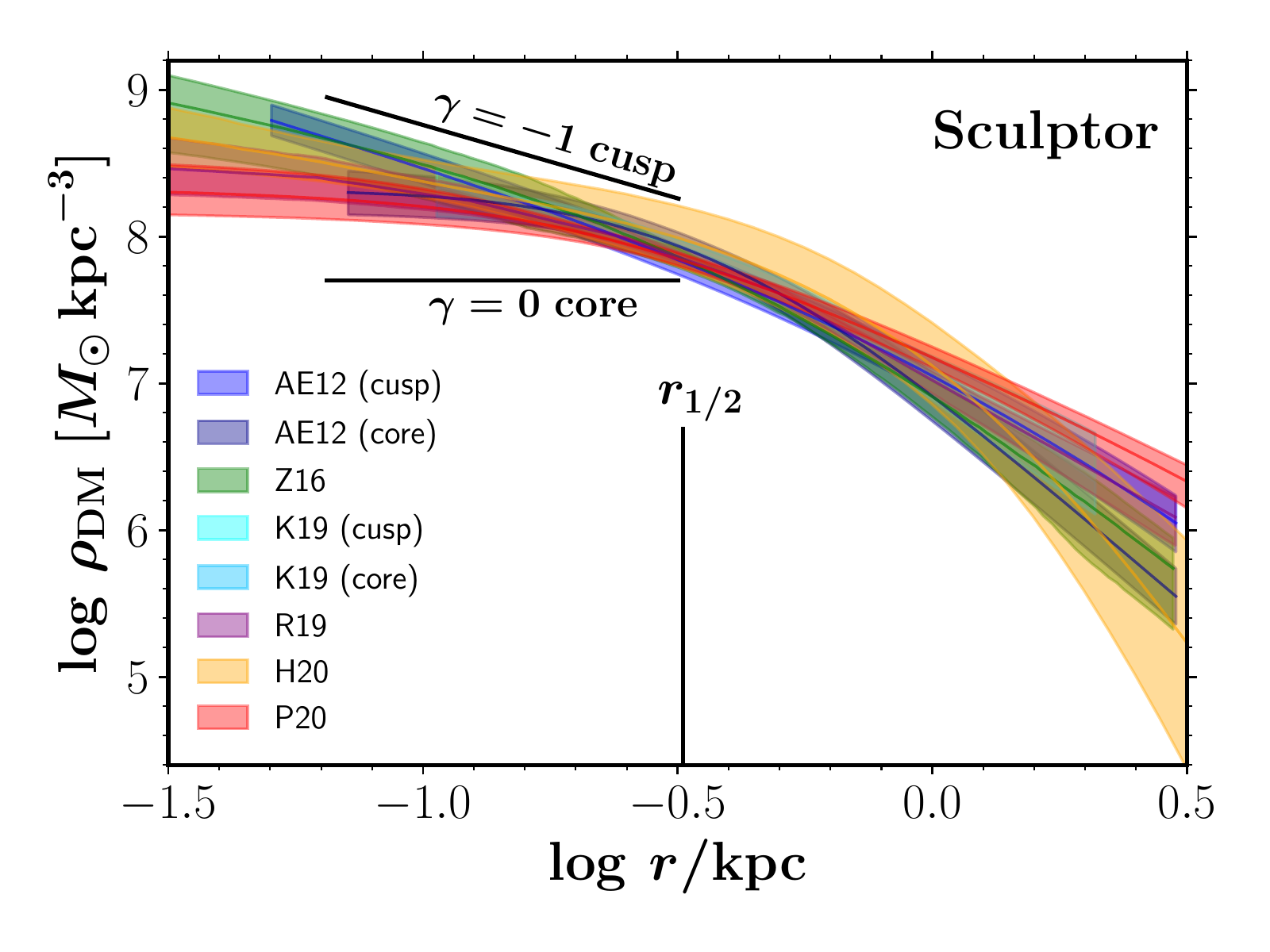}
\caption{Same as Fig.~\ref{fig:draco}, but for the Sculptor dSph. Here the profiles are taken from Amorisco \& Evans (AE12, \cite{Amorisco2012}; both cuspy and cored models), 
Zhu et al.\ (Z16, \cite{Zhu2016}), Kaplinghat et al.\ (K19, \cite{Kaplinghat2019}; both cuspy and cored models), Read et al.\ (R19, \cite{Read2019}), Hayashi et al.\ (H20, \cite{Hayashi2020}; circularized spheroidal model) and Pascale (P20, \cite{Pascale2020}). The $\Mstar$ profile is taken from Read et al.\ \citep{Read2019}.  }
\label{fig:sculptor}
\end{figure}

Knowing the relative distribution of DM and baryonic mass in DGs is a fundamental piece of the puzzle of DG formation and evolution (see, e.g.,  section 10.9 of \cite{Cimatti2019}). 
The Draco, Fornax and Sculptor dSphs, for which rich photometric and spectroscopic datasets are available, are interesting to look at in some detail, because they are illustrative of the variety of DM halo properties inferred for LG DGs with spatially resolved information on the mass distribution. A representative selection of  $\Mdyn(r)$ profiles of these three classical dSphs obtained in a few literature works are shown in the upper panels of Figs.~\ref{fig:draco}-\ref{fig:sculptor}, together with the $\Mstar(r)$ profile as estimated by \cite{Read2019}.
The mass profiles are such that  $\Mdyn(r)$ is significantly higher than $\Mstar(r)$ down to the innermost probed radii.
It is clear that $\Mdyn$ is best constrained at radii slightly larger than the stellar half-mass radius $\rhalf$, where the effect of the mass-anisotropy degeneracy is minimum (see Section~\ref{sec:dynmod}). 

The lower panels of Figs.~\ref{fig:draco}-\ref{fig:sculptor} show, for the same selection of works as in the corresponding upper panels, the DM density profiles of Draco, Fornax and Sculptor.
The inner logarithmic slope $\gamma\equiv \d\ln \rhoDM / \d \ln r$ of the halo density profile is a relevant quantity in the context of cosmological and galaxy formation models \citep[e.g.][]{Bullock2017}. 
In particular it is interesting to determine whether the dark halos of dwarfs are cuspy ($\gamma\approx -1$), as found in halos formed in DM-only cosmological simulations  within the standard CDM paradigm \citep[][]{Navarro1996b}, or cored ($\gamma\approx 0$), which might be a signature of baryon physics and star formation within CDM halos \citep[][]{Navarro1996a,Read2005,Mashchenko2008,DiCintio2014,Madau2014,Nipoti2015,Sawala2016,Benitez-Llambay2019} or of physics of DM different from standard CDM (e.g.\ \cite{Bullock2017,Hui2017,Fitts2019,Burger2021}).
In the lower panels of Figs.~\ref{fig:draco}-\ref{fig:sculptor}  power laws 
$\rhoDM\propto r^{\gamma}$ with $\gamma=-1$ (cusped density profile) and
$\gamma=0$ (cored density profile) are shown over the radial range $0.2\rhalf<r<\rhalf$, taken as representative of the inner parts of the halo. 

In Draco the stellar mass is negligible at all radii (see upper panel of Fig.~\ref{fig:draco}): $\Mdyn$ is a factor of $\approx 100$ higher than $\Mstar$ within $\rhalf\approx0.25\kpc$, and a factor of $\approx 1000$  within $2\kpc$. Draco is believed to have a cuspy DM halo \citep{Jardel2013b,Read2018,Hayashi2020}, as also suggested by the DM density profiles shown in bottom panel of Fig.~\ref{fig:draco}. 
In  Fornax  $\Mdyn$ is almost a factor of ten higher than $\Mstar$ within $\rhalf\approx0.85\kpc$, and the trend is that $\Mdyn/\Mstar$ gradually increases outwards (Fig.~\ref{fig:fornax}, upper panel). 
Both cuspy and cored DM halo models have been proposed in the literature for Fornax (see bottom panel Fig.~\ref{fig:fornax}): when dynamical models are systematically compared with the data, in some studies cored density profiles are favoured \citep[][]{Jardel2012,Pascale2018,Read2019,Hayashi2020}, while in others cuspy and cored models are found to be perform similarly \citep[][]{Breddels2013a,Kaplinghat2019}.
The properties of Sculptor are intermediate between those of Fornax and those of Draco: 
$\Mdyn$ is more than a factor of ten higher than $\Mstar$ within $\rhalf\approx 0.32\kpc$, and up to a factor $\approx 100$ higher within $2\kpc$ (Fig.~\ref{fig:sculptor}, upper panel). It is highly debated whether the DM halo of Sculptor is cuspy or cored (e.g.\ \cite{Amorisco2012,Breddels2013b,Strigari2017,Strigari2018a,Pascale2020}; see bottom panel of Fig.~\ref{fig:sculptor}), which is possibly due to the fact that the cusp is mild \citep[][]{Zhu2016,Read2019,Hayashi2020}.

A possible trend, well represented by Fornax, Draco and Sculptor, is that DM cusps tend to be present in DGs that are highly DM dominated and with no recent star formation (such as Draco), while DGs with non-negligible stellar mass and more continuous star formation history (such as Fornax) tend to have cored halos \citep[][]{Read2019,Hayashi2020}, a picture supported also by the results of some hydrodynamic cosmological simulations of DG formation \citep[e.g.][]{Pontzen2014,Tollet2016,Lazar2020b}. MW dSphs with extended star formation histories are however rare. LG dIrrs and dTs would be a natural testing ground to assess the impact of star formation and stellar feedback, but in general the available data do not allow us to perform detailed dynamical modelling. An exception is the dIrr WLM for which a joint analysis of stellar and gas kinematics has provided evidence that the halo is cored and prolate \cite{Leung2021}.

When discussing whether DM halos are cored or cuspy, one must bear in mind that the density profiles of DM halos, even in the central regions, are not necessarily well approximated by power laws, so the logarithmic density slope $\gamma(r)$ can vary significantly with radius (see e.g.\ section 7.5 of \cite{Cimatti2019}). Some authors provide measurements of $\gamma$ at a given physical radius, but another possible choice is to measure $\gamma$ at a given fraction of the stellar-half mass radius $\rhalf$, which can be very different, in physical units, in different DGs. In any case it is important to make coherent choices when comparing, for instance, simulations and observations (see \cite{Battaglia2013}).

The DM density
profile is characterised not only by its radial trend 
but also by its normalisation,  so an important complementary
information is the value of the DM density at or within a reference radius, either in physical units (such as $300\pc$ as in \cite{Strigari2008} or  $150\pc$ as in \cite{Read2019}) or as a fraction of $\rhalf$. 
When looking at LG DGs as a population,  lower-luminosity galaxies tend to have centrally denser DM profiles \citep[][]{Read2019,Hayashi2020}. This could be interpreted as an indication that the dwarf-size halos in which star formation has been extremely inefficient have maintained a profile more similar to that of the pristine cosmological halos. However, this could be only part of the story, because
there are indications that,
for the classical MW dSphs, $\rhoDM(150\pc)$ anticorrelates with the pericentric radius of the dwarf's orbit \citep[][]{Kaplinghat2019,Hayashi2020},
which suggests that also tidal interactions can have a role \citep[see][]{Robles2021}. A different case is that of the UFD Crater~II \citep[][]{Torrealba2016,Caldwell2017}, which is believed to have suffered strong tidal interactions, but has very low DM density \citep[][]{Fu2019,Borukhovetskaya2021b,Ji2021}. Further studies are necessary to fully understand the role of tidal fields in shaping the DM halos of DGs. 

Estimates of the DM density distribution of LG DGs are particularly interesting in the context of indirect DM detection experiments via observation of $\gamma$ rays produced by annihilation or decay of  DM particles.
For this purpose it is useful to provide as output of the dynamical modelling the so-called annihilation $J$-factor
$J\propto \rhoDM^2$ and decay $D$-factor $D\propto\rhoDM$ (e.g.\ \cite{Evans2016}), which are the astrophysical factors to which, respectively, the annihilation and decay fluxes are proportional for given DM particle model. 
By definition, $J$ and $D$ depend not only on the intrinsic DM distribution, but also on the distance $d$ of the DG:
\cite{Pace2019} find for LG DGs the empirical scaling laws $J\propto\sigmalos^4d^{-2}\Rhalf^{-1}$ and $D\propto\sigmalos^2d^{-2}\Rhalf$.
From the point of view of indirect DM detection, the most interesting DGs are those with the highest  values of the $J$ and $D$ factors \citep[e.g.][]{Bergstrom2018,Strigari2018b}.
The closest UFDs tend to have the highest estimated  J and D factors \citep[][]{Ackermann2015,Bonnivard2015,Geringer-Sameth2015,Evans2016,Sanders2016b,Pace2019}, though with relatively large associated uncertainties.  Also some of the classical dwarfs, such as Draco, Sculptor and Ursa Minor, have promisingly high J and D factors \citep[][]{Bonnivard2015,Hayashi2016,Klop2017,Pascale2018,Chiappo2019,Horigome2020}, with the advantage, over UFDs, that the estimates are more robust thanks to the richer kinematic samples.

\section{Outlook} \label{sec:outlook}

In the past decade, there have been several advances in our observational understanding of stellar kinematics and DM content of LG DGs. 

Half of the known LG DGs, in particular the faintest or most diffuse systems around the MW, were discovered and characterised. Apart from alleviating the missing satellite problem (e.g.\ \cite{Bullock2017}), these extreme systems give particularly interesting insights into the process of galaxy formation, and its stochasticity, in low mass DM halos. UFDs are the most DM dominated galaxies and are thus in principle prime targets for DM studies.
However, only integrated quantities such as the DM mass within a given radius can currently be determined for most UFDs. Inferring their mass distribution necessitates significantly enlarging the spectroscopic samples by reaching much dimmer magnitudes than what possible with 10m-class telescopes; for this, we might have to await the multi-object spectrographs with field-of-views of a few arcmin on 30m-40m class telescopes. 
An open question for UFDs is how much their measured $\sigmalos$ might be inflated by unresolved binaries (see Box~1). So far, the great majority of studies determining the properties of the binary star populations in LG DGs relied on data-sets not specifically built for this purpose, e.g.\ with  fairly large l.o.s.\ velocity uncertainties,
relatively short time samplings and heterogeneous data-sets. Observing programs specifically designed for this goal would be of value, especially if also constraints on the period distribution can be provided.

Even though the data-sets are still not at a similar level as those for the nearby MW classical dSphs, significant efforts were made towards an improved quantification of the properties of isolated LG DGs and M31 satellites. 
Studying these systems gives access to a larger variety of star formation histories than those exhibited by MW satellites, opening the road to useful applications for understanding the impact of star formation and stellar feedback on DM halo properties.

Where available, large and spatially extended spectroscopic samples with precise velocities and metallicity/age information have often unveiled the presence of multiple stellar components. It would be desirable to gather further data-sets that would allow to establish the presence and properties of such components in a larger number of LG DGs, because their simultaneous dynamical modelling can effectively alleviate the mass-anisotropy degeneracy.
In addition, it would be important to pin down which systems have their stellar kinematics free from “peculiarities", or at least to understand to what extent these might impact inferences of the DM halo properties. 

Highly multiplexed fiber-fed spectrographs with fields of view from half to a few degrees diameters, (to be) mounted on 4m-10m class telescopes, such as DESI at the KPNO Mayall Telescope, WHT/WEAVE, VISTA/4MOST, VLT/MOONS, Subaru/PFS, CFHT/MSE will be the protagonists in the next generation of explorations of the properties of bright LG DGs, possibly coupled to instruments apt to probing crowded regions, such as VLT/MUSE. 

Exciting advances were made thanks to the ESA {\it Gaia} mission, which has allowed a factor of $\sim$7 increase in the number of LG DGs with a measurement of the transverse bulk motion \citep{Battaglia2021}, and, in combination with HST, enabled resolving the internal random motions in the proper motion component for two MW dSphs. In the future, imaging assisted by Adaptive Optics on the Extremely Large Telescopes (ELTs) could potentially play an important role in assembling the proper motion measurements needed to tackle the inner slope of MW dSphs DM halo density profile  \citep[e.g.][]{Simon2019_WP}. Nonetheless, the fields-of-view of HST and ELT instruments encompass only a few percent of the stellar body of MW dSphs and it would certainly be beneficial to cover much larger areas to the aim of obtaining velocity dispersion {\it profiles} also in the proper motion components. This is one of the tantalizing possibilities that the combination of data from {\it Gaia} and Gaia-NIR \cite{Hobbs2021} or the Roman Space Telescope \cite{WFIRST2019} opens. 
Even though precision astrometry with ground-based imagers is certainly challenging, it is actually possible to pursue  \citep[e.g.][]{Anderson2006} and it might be worth exploring whether the right observing strategy could make it within the reach of the Vera Rubin Observatory to resolve internal proper motions dispersions over a large area for a couple of MW dSphs. 

As far as dynamical modelling of DGs is concerned, we have seen that different degrees of complexity are possible both in the realism of the models and in the kind of input data used.  The theorists, relying on tests with mock data (e.g.\ \citep{Read2021}), can on one hand tune the sophistication of the adopted model, depending on the quality and quantity of available data, and, on the other hand, guide the observers in acquiring the data-sets. When the data expected from the aforementioned future facilities are available, it will become more and more important to adopt non spherical DF- or orbit-based models, with multiple stellar populations, and to use individual star measurements or high velocity moments as input data. 
The main aim of these efforts will be to obtain precise measurements of the density distributions of the DM halos of LG DGs, with the twofold objective of trying to unveil the nature of DM and improving our knowledge of the process of galaxy formation. 

But what is the clear path to success to shed light onto the nature of DM from dynamical modelling of LG DGs? In our opinion, there is no easy answer - yet - to this question. 
Up to few years ago, there was a well defined goal: searching for the DM cusps predicted by CDM-only cosmological simulations. The inclusion of baryonic physics in the simulations, including star formation and stellar feedback, has painted a much more complex and still controversial picture. There is no univocal prediction on the impact of these processes on the inner density profile of an initially cuspy DM halo.
Presently, we are therefore in the situation in which the best foreseeable contribution of dynamical modelling of LG DGs to the field is to determine as accurately as possible the DM halo properties of a variety of systems, for instance differing significantly in baryon fraction or star formation history. 

Currently, there are still on-going efforts aimed at understanding what precision can be achieved in the various parameters describing the DM halo density distribution as a function of sample size and different combination of observables, in lack of the full 6D information \citep[e.g.][]{Strigari2007,Richardson2014planb,Read2017b,Guerra2021}, as well as comparing the precision and accuracy of various modelling techniques \citep[e.g.][]{Read2021}. The aforementioned studies show that pushing l.o.s.\ velocities to better than $\sim$5 km s$^{-1}$ precision does not appear to yield significant gains, while  noticeable improvements are brought by the inclusion of proper motion measurements or distances to the individual stars, in particular if the samples exceed 1000 stars.  
We have seen that, on the observational front,  the future looks bright, with no shortage of possibilities for gathering exquisite data-sets. Naturally, such endeavours would strongly benefit by guidance from theoretical models in identifying what observational tests 
would be required in order to distinguish between possible competing models.

\backmatter

\bmhead{Acknowledgments}

We are grateful to Nicola Amorisco,  Kohei Hayashi, Manoj Kaplinghat, Raffaele Pascale, Justin Read, Marco Valli and Ling Zhu for sharing their data. GB acknowledges support from the State Research Agency (AEI), Spanish Ministry of Science, Innovation and Universities (MICIU), the Agencia Estatal de Investigación del Ministerio de Ciencia e Innovación(AEI-MCINN) and the European Regional Development Fund (ERDF) under grants with reference  AYA2017-89076-P (AEI/FEDER, UE) and PID2020-118778GB-I00, and the “Centro de Excelencia Severo Ochoa” program through grants no. CEX2019-000920-S, as well as by the MICIU, through to the State Budget and by the Canary Islands Department of Economy, Knowledge and Employment, through the Regional Budget of the Autonomous Community.

\begin{glos}[Glossary]

\item[CDM] Cold dark matter

\item[CKC] Chemo-kinematic component

\item[DF] Distribution function

\item[DG] Dwarf galaxy

\item[dIrr] Dwarf irregular galaxy

\item[DM] Dark matter

\item[dSph] Dwarf spheroidal galaxy

\item[dT] Dwarf galaxy of transition type

\item[LG] Local Group 

\item[MW] Milky Way

\item[RAR] Radial acceleration relation 

\item[RGB] Red giant branch

\item[UFD] Ultrafaint dwarf

\end{glos}

\end{document}